\documentclass[preprintnumbers,twocolumn,prd,aps,superscriptaddress,footinbib,amsfonts,amsmath,amssymb,showpacs,floatfix]{revtex4-1}

\usepackage[utf8]{inputenc}
\usepackage{amsmath,amssymb,amsfonts,mathrsfs,bbold}
\usepackage{yfonts}
\usepackage{graphicx}
\usepackage[colorlinks=true,citecolor=blue,linkcolor=blue,urlcolor=blue]{hyperref}
\usepackage[dvipsnames]{xcolor}
\usepackage{lipsum}
\usepackage{slashed}
\usepackage[normalem]{ulem}
\usepackage{url}

\usepackage{amssymb, bm, amsmath, graphicx}
\usepackage{enumerate}
\usepackage{amsfonts}
\usepackage{mathtools}
\usepackage{latexsym}
\usepackage{epstopdf}

\def\eq#1{{Eq.~(\ref{#1})}}
\def\fig#1{{Fig.~\ref{#1}}}
\newcommand{\ben}{\begin{eqnarray*}}
\newcommand{\een}{\end{eqnarray*}}

\newcommand{\as}{\alpha_s}

\preprint{JLAB-THY-21-3318}

\begin{document}

\preprint{JLAB-THY-21-3318}

\title{First analysis of world polarized DIS data with small-$x$ helicity evolution}

\author{Daniel Adamiak }
\email[Email: ]{adamiak.5@osu.edu}
\affiliation{Department of Physics, The Ohio State University, Columbus, Ohio 43210, USA}
\author{Yuri V. Kovchegov}
\email[Email: ]{kovchegov.1@osu.edu}
\affiliation{Department of Physics, The Ohio State University, Columbus, Ohio 43210, USA}
\author{W. Melnitchouk}
\affiliation{Jefferson Lab, Newport News, Virginia 23606, USA}
\author{Daniel Pitonyak }
\email[Email: ]{pitonyak@lvc.edu}
\affiliation{Department of Physics, Lebanon Valley College, Annville, Pennsylvania 17003, USA}
\author{Nobuo Sato}
\email[Email: ]{nsato@jlab.org}
\affiliation{Jefferson Lab, Newport News, Virginia 23606, USA}
\author{Matthew D. Sievert}
\email[Email: ]{msievert@nmsu.edu}
\affiliation{Department of Physics, New Mexico State University, Las Cruces, New Mexico 88003, USA \\
       \vspace*{0.2cm}
       {\bf Jefferson Lab Angular Momentum (JAM) Collaboration
        \vspace*{0.2cm}}}



\begin{abstract}
We present a Monte Carlo based analysis of the combined world data on polarized lepton-nucleon deep-inelastic scattering at small Bjorken $x$ within the polarized quark dipole formalism.
We show for the first time that double-spin asymmetries at $x<0.1$ can be successfully described using only small-$x$ evolution derived from first-principles QCD, allowing predictions to be made for the $g_1$ structure function at much smaller $x$.
Anticipating future data from the Electron-Ion Collider, we assess the impact of electromagnetic and parity-violating polarization asymmetries on $g_1$ and demonstrate an extraction of the individual flavor helicity PDFs at small $x$. 

\end{abstract}

\date{\today}
\maketitle

 \section{Introduction}
The partonic origin of the proton spin remains one of the most intriguing and persistent problems in hadronic physics.
Spin sum rules \cite{Jaffe:1989jz, Ji:1996ek} 
decompose the proton spin of $1/2$ (in units of $\hbar$) into the contributions from quark and gluon helicities ($\Delta\Sigma$, $\Delta G$) and orbital angular momenta.
Extensive experimental programs at facilities around the world over the past three decades have provided important insights into the proton spin decomposition \cite{Aidala:2012mv}.  However, outstanding questions remain, especially about the detailed momentum dependence of the associated quark and gluon helicity parton distribution functions (PDFs) $\Delta q$ and $\Delta g$, respectively.
These PDFs are related to the total quark and gluon spin contributions to the proton spin via integrals
over the partonic momentum fraction~$x$,
\begin{subequations}
\label{eqn:moments}
\begin{align}
& \Delta\Sigma(Q^2) = \sum_q \int_0^1 dx \; \Delta q^+(x,Q^2)\,,
\label{eqn:Sq} \\
& \Delta G(Q^2) = \int_0^1 dx \; \Delta g(x,Q^2)\,,
\label{eqn:SG}
\end{align}
\end{subequations}
where $\Delta q^+ \equiv \Delta q + \Delta {\bar q}$, and the sum runs over the quark flavors $q=u$, $d$, $s$, with $Q^2$ the resolution scale.

Determining the quark and gluon contributions to the proton spin crucially depends on knowing the $x$ dependence of the PDFs $\Delta q^+(x,Q^2)$ and $\Delta g(x,Q^2)$.  This is especially true at small values of $x$, where the computation of the moments (\ref{eqn:moments}) involves extrapolation below the experimentally accessible region, down to $x=0$.
In recent years, an effort to develop small-$x$ evolution equations for helicity PDFs has been underway~\cite{Kovchegov:2015pbl, Hatta:2016aoc, Kovchegov:2016zex, Kovchegov:2016weo, Chirilli:2018kkw, Kovchegov:2019rrz, Boussarie:2019icw,Chirilli:2021lif}, building in part on Refs.~\cite{Kirschner:1983di, Bartels:1995iu, Bartels:1996wc}. 
Specifically, small-$x$ evolution equations (herein referred to as KPS evolution) for the so-called ``polarized dipole amplitude'' have been derived~\cite{Kovchegov:2015pbl, Kovchegov:2016zex, Kovchegov:2016weo, Kovchegov:2017jxc, Kovchegov:2017lsr, Kovchegov:2018znm, Cougoulic:2019aja}.

The polarized dipole amplitude is a critical object for spin-dependent phenomena at small values of $x$~(see~Fig.~\ref{fig:DIS}):~it allows one to obtain the spin-dependent $g_1$ structure function, along with the (collinear and transverse momentum dependent) helicity PDFs~\cite{Kovchegov:2015pbl, Kovchegov:2016zex}.
At leading order (LO) in the strong coupling $\alpha_s$, these equations resum powers of $\as \, \ln^2 (1/x)$, which is known as the double-logarithmic approximation (DLA).
The KPS evolution equations close in the large-$N_c$ limit~\cite{Kovchegov:2015pbl}, where $N_c$ is the number of colors.  Numerical and analytic solutions for these have previously been constructed~\cite{Kovchegov:2016weo, Kovchegov:2017jxc, Kovchegov:2017lsr}.
However, an analysis of the world polarized deep-inelastic scattering (DIS) data at small $x$ utilizing KPS evolution has never been performed.

In this Letter, we present such an analysis.  We emphasize that KPS evolves in $x$ instead of the traditional evolution in $Q^2$~\cite{Gribov:1972ri, Altarelli:1977zs, Dokshitzer:1977sg}.
Unpolarized small-$x$ evolution~\cite{Balitsky:1995ub, Balitsky:1998ya, Kovchegov:1999yj, Kovchegov:1999ua, Kovchegov:2006vj, Balitsky:2006wa} was previously used to describe DIS data on the proton $F_2$ and $F_L$ structure functions~\cite{Albacete:2010sy, Albacete:2009fh, Beuf:2020dxl}. 
We show for the first time that an analogous helicity-dependent small-$x$ approach can successfully describe the polarized DIS $g_1$ structure function for the proton and neutron extracted from data at $x<0.1$.
This approach differs from earlier work~\cite{Blumlein:1996hb} which incorporated the small-$x$ resummation from Ref.~\cite{Bartels:1996wc} into the polarized DGLAP splitting functions~\cite{Gribov:1972ri, Altarelli:1977zs, Dokshitzer:1977sg}, thereby mixing the small-$x$ and $Q^2$ resummations.

In addition, we use pseudodata from the future Electron-Ion Collider (EIC) on electromagnetic and parity-violating polarization asymmetries to demonstrate an extraction of helicity PDFs at small~$x$ within the KPS formalism and assess the impact on $g_1$.
This is a first step towards ultimately using small-$x$ evolution with experimental data from various reactions to genuinely {\it predict} the amount of spin carried by small-$x$ partons, which is crucial to resolving the puzzle of the partonic origin of the proton spin.

\section{Formalism} 
\begin{figure}
\centering
\includegraphics[width=0.42 \textwidth]{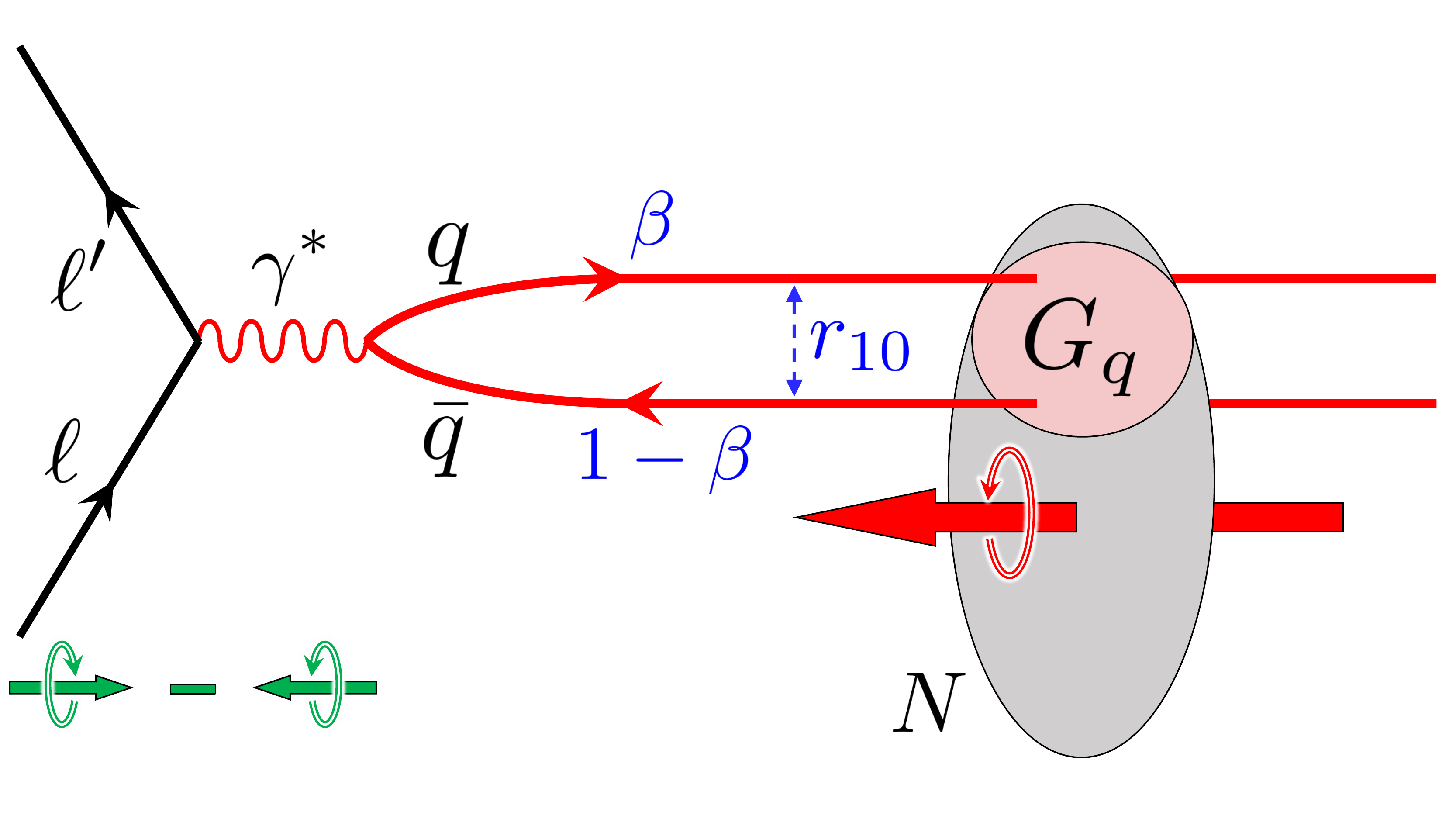}
\caption{Illustration of polarized DIS at small $x$.  The exchanged virtual photon fluctuates into a $q \bar{q}$ dipole of transverse size $r_{10}$, with $\beta$ the fractional energy carried by the less energetic parton in the dipole.  The spin-dependent scattering amplitude of the dipole on the polarized nucleon $N$ is described by $G_q(r_{10}^2, \beta s)$, producing an asymmetry between the cross sections for positive and negative helicity leptons.}
\label{fig:DIS}
\end{figure}

In the DLA the quark helicity PDFs can be written in terms of the polarized dipole amplitude $G_q\big(r_{10}^2, \beta s\big)$~\cite{Kovchegov:2015pbl, Kovchegov:2016zex, Kovchegov:2016weo} (see \fig{fig:DIS}),
\begin{align}
\label{Dq+1}
    \Delta q^+(x,Q^2) = \frac{N_c}{2 \pi^3} 
    \int\limits_{\Lambda^2/s}^1 \frac{d\beta}{\beta} 
    \int\limits_{1/\beta s}^{r^2_{\rm max}} \frac{d r_{10}^2}{r_{10}^2}\, 
    G_q\big(r_{10}^2, \beta s\big),
\end{align}
where $s \approx Q^2 (1-x)/x$ is the invariant mass squared of the $\gamma^* N$ system and $\beta$ is the fraction of the virtual photon's momentum carried by the less energetic parton in the $q \bar{q}$ dipole. 
The amplitude $G_q$ is also integrated over all impact parameters~\cite{Kovchegov:2015pbl, Kovchegov:2016zex, Kovchegov:2016weo, Kovchegov:2017jxc, Kovchegov:2017lsr, Kovchegov:2018znm, Cougoulic:2019aja}, 
    $r_{10} = |\bm{r}_1 - \bm{r}_0|$ 
is the dipole transverse size, where 
$\bm{r}_i$ is a coordinate vector in the transverse plane, and  
      $r^2_{\rm max} = \min \!\left\{ 1/\Lambda^2, 1/(\beta Q^2) \right\}$.
We regulate the long-distance behavior of $r_{10}$ with an infrared cutoff $1/\Lambda$ and set $\Lambda=1\,{\rm GeV}$.

Changing variables to 
\begin{align}
    \eta = \sqrt{\frac{\alpha_s N_c}{2\pi}} \ln \frac{\beta s}{\Lambda^2}, \ \ \ \ s_{10} = \sqrt{\frac{\alpha_s N_c}{2\pi}} \ln \frac{1}{r_{10}^2 \Lambda^2},
\end{align}
we can rewrite \eq{Dq+1} in the form~\cite{Kovchegov:2016weo}
\begin{align}\label{Dq+3}
    \Delta q^+(x,Q^2) = \frac{1}{\alpha_s \pi^2}
    \int\limits_0^{\eta_{\rm max}} d\eta 
    \int\limits_{s_{10}^{\rm min}}^\eta ds_{10}\, G_q\big(s_{10}, \eta\big),
\end{align}
where the limits on the $\eta$ and $s_{10}$ integrations are given by
    $\eta_{\rm max} = \sqrt{\alpha_s N_c/2\pi}\, \ln(Q^2/x\Lambda^2)$,
and
    $s_{10}^{\rm min} = \max \!\left\{ \eta - \sqrt{\alpha_s N_c/2\pi} \ln(1/x), 0 \right\}$,
respectively.

In the large-$N_c$ limit
the polarized dipole amplitude $G_q$ obeys the evolution equations~\cite{Kovchegov:2015pbl, Kovchegov:2016zex, Kovchegov:2016weo},
\begin{subequations} \label{e:HelEv2_f}
\begin{align}
    &G_q(s_{10}, \eta) = G_q^{(0)} (s_{10}, \eta)  \\
    & \hspace*{0.5cm}
    + \int\limits_{s_{10}}^{\eta} d\eta' 
      \int\limits_{s_{10}}^{\eta'} d s_{21} \: 
      \big[ \Gamma_q (s_{10} , s_{21} , \eta') + 3\, G_q (s_{21}, \eta') \big] , \nonumber\\
    &\Gamma_q (s_{10} , s_{21} , \eta') = G_q^{(0)}(s_{10} , \eta') \\ 
    &\hspace*{0.5cm} 
    + \int\limits_{s_{10}}^{\eta'} d\eta''
     \int\limits_{s_{32}^{\rm min}}^{\eta''} ds_{32} \, 
     \big[ \Gamma_q (s_{10} , s_{32} , \eta'') + 3\, G_q (s_{32}, \eta'') \big], \nonumber
\end{align}
\end{subequations}
where $s_{32}^{\rm min}\! =\! \max\!\left\{\!s_{10} \, , \, s_{21} - \eta' + \eta''\right\}$, and $\Gamma_q(s_{10},s_{21},\eta')$ is an auxiliary polarized ``neighbor'' dipole amplitude, defined in Ref.~\cite{Kovchegov:2015pbl}, whose evolution mixes with $G_q\!\left(s_{10}, \eta \right)$.
Note that only $G_q\!\left(s_{10}, \eta \right)$ contributes to $\Delta q^+$ in \eq{Dq+3}.
The evolution kernel in Eqs.~\eqref{e:HelEv2_f} is LO in $\alpha_s$ and has been further simplified to contain only the DLA terms.
Since running coupling corrections are higher order, we freeze the coupling in Eq.~\eqref{Dq+3} at $\alpha_s = 0.3$, a typical value in the DIS $Q^2$ range we study.

For given initial conditions  $G_q^{(0)}(s_{10}, \eta)$, we can solve Eqs.~\eqref{e:HelEv2_f} for $G_q (s_{10}, \eta)$ and use it in \eq{Dq+3} to calculate $\Delta q^+$.
Inspired by the Born-level perturbative calculation of $G_q(s_{10}, \eta)$ \cite{Kovchegov:2015pbl, Kovchegov:2016zex, Kovchegov:2016weo}, we employ the {\it ansatz}
\begin{align}\label{G01}
    G_q^{(0)}(s_{10}, \eta) = a_q \, \eta + b_q \, s_{10} + c_q
\end{align}
for the initial conditions, with flavor-dependent coefficients $a_q$, $b_q$, and $c_q$ ($q = u, d, s$) as free parameters.

The evolution in Eqs.~\eqref{e:HelEv2_f} starts at $\eta = s_{10}$, or $\beta s = 1/ r_{10}^2$. 
Since $r_{10} \sim 1/Q$ and the $\beta$ integral in \eq{Dq+1} extends up to 1, the evolution in Eqs.~\eqref{e:HelEv2_f} begins at $x=1$. 
This cannot be the case for small-$x$ evolution, so \eqref{e:HelEv2_f} must be modified to reflect the start of evolution only at $x = x_0 \ll 1$.
For unpolarized small-$x$ evolution, which can be written as a differential equation in $x$, this usually means that one only needs to set the initial conditions at $x = x_0$~\cite{Albacete:2010sy, Albacete:2009fh, Beuf:2020dxl}.
However, the modifications in the polarized case are more involved because \eqref{e:HelEv2_f} are integral equations and cannot be cast in a differential form. 
Defining $y_0 \equiv \sqrt{\alpha_s N_c/2\pi}\ln(1/x_0)$, for $\eta - s_{10} > y_0$ and $\eta' - s_{10} > y_0$, the modified evolution equations are
\begin{subequations} \label{e:HelEv3_f}
\begin{align}
    &G_q (s_{10}, \eta) = G_q^{(0)}(s_{10}, \eta)  \\
    &+ 
    \int\limits_{s_{10}+y_0}^{\eta} d\eta' 
    \int\limits_{s_{10}}^{\eta' - y_0} d s_{21} \: 
    \big[ \Gamma_q (s_{10} , s_{21} , \eta') + 3\, G_q (s_{21}, \eta') \big] , \nonumber\\
    &\Gamma_q (s_{10} , s_{21} , \eta') = G_q^{(0)}(s_{10} , \eta')  \\ 
    &+ \int\limits_{s_{10}+y_0}^{\eta'} d\eta''
    \int\limits_{s_{32}^{\rm min}}^{\eta'' - y_0} ds_{32} \, 
    \big[ \Gamma_q (s_{10} , s_{32} , \eta'') + 3\, G_q (s_{32}, \eta'') \big] . \nonumber
\end{align}
\end{subequations}
In the region below $y_0$, the polarized dipole amplitude is given by the initial conditions     $G_q(s_{10},\eta-s_{10}<y_0) 
    = \Gamma_q(s_{10},s_{21},\eta'-s_{10}<y_0) 
    = G_q^{(0)}(s_{10},\eta)$. 
This prescription implements our matching onto large-$x$ physics, with development of a more rigorous matching procedure left for future work. The numerical solution of Eqs.~\eqref{e:HelEv3_f} is accomplished with the discretization utilized in  Ref.~\cite{Kovchegov:2016weo} and employing the algorithm presented in Ref.~\cite{Kovchegov:2020hgb}.

\section{Observables}
In this work we focus on polarized inclusive DIS data to demonstrate that KPS evolution can describe the existing measurements at small $x$ using the simple initial conditions \eqref{G01}.
The main observables used in our analysis are the double-longitudinal spin asymmetries $A_{||}$ and $A_1$ from the scattering of polarized leptons on polarized nucleons.
At large $Q^2$, these are given by ratios of the $g_1$ to $F_1$ structure functions, $A_{||} \propto A_1 \propto g_1/F_1$,
where in the DLA the $g_1$ structure function is
\begin{align}
    & g_1(x,Q^2) = \frac{1}{2}\sum_q e_q^2\, \Delta q^+(x,Q^2). 
    \label{e:g1LO}
\end{align}
The denominator $F_1$ is taken from data in the form of the LO JAM global analysis~\cite{Cocuzza21,Sato:2019yez}. Note that to this order the Bjorken $x$ variable coincides with the partonic momentum fraction, although at higher orders these are of course different.

Analyses solely utilizing  inclusive proton and neutron (deuteron or $^3$He) DIS data~\cite{Jimenez-Delgado:2013boa, Sato:2016tuz} 
need additional input 
to separately determine each of the flavors 
$\Delta u^+$, $\Delta d^+$, and $\Delta s^+$. 
This can be partially achieved by assuming SU(3) flavor symmetry in the sea and employing the octet axial charge,
    $a_8 = \int_0^1 \!dx \big( \Delta u^+ + \Delta d^+ - 2\Delta s^+ \big)$, 
as a constraint on these moments. However, this is insufficient to uniquely determine the $x$ dependence, so at least one more observable is needed to solve for all three distributions.
One approach is to include semi-inclusive DIS (SIDIS) data, with $\pi$ and $K$ fragmentation functions (FFs) as tags of individual flavors. However, to avoid additional uncertainties due to FFs, which would need to be fitted simultaneously with the PDFs~\cite{Ethier:2017zbq, Sato:2019yez, Moffat:2021dji}, we leave this to future work.

A new opportunity presented by the future EIC, in addition to precision measurements of $A_{||}$ at smaller values of $x$, is the possibility to perform parity-violating (PV) DIS with unpolarized electrons scattering from longitudinally polarized nucleons.
By utilizing the interference between the electromagnetic and weak neutral currents, the resulting asymmetry $A_{\rm PV}$ can provide independent combinations of helicity PDFs that could allow clean flavor separation at low~$x$.

One contribution to the $A_{\rm PV}$ asymmetry comes from the lepton axial vector--hadron vector coupling, which is proportional to the $g_1^{\gamma Z}$ interference structure function, weighted by the weak axial vector electron charge $g_A^e = -\frac12$.  The other comes from the lepton vector--hadron axial vector coupling, given by the $g_5^{\gamma Z}$ structure function weighted by the weak vector electron charge, $g_V^e = -\frac12 (1-4 \sin^2\theta_W)$~\cite{Hobbs:2008mm, Zhao:2016rfu}.
The $g_5^{\gamma Z}$ structure function provides information on nonsinglet combinations $\Delta q^- \equiv \Delta q - \Delta\bar{q}$. However, since  $|g_V^e| \ll 1$, and at small~$x$ one has $\Delta q^- \ll \Delta q^+$~\cite{Kovchegov:2016zex}, its contribution to $A_{\rm PV}$ is strongly suppressed.
For three quark flavors, the PV asymmetry is then determined by the ratio $g_1^{\gamma Z}/F_1$, where in the DLA we have, 
\begin{align}
    & g_1^{\gamma Z} (x, Q^2) = \sum_q e_q\, g_V^q\, \Delta q^+(x, Q^2),
\label{e:g1gamZ} 
\end{align}
with $g_V^q = \pm\frac12 - 2 e_q \sin^2\theta_W$ the weak vector coupling to $u$- and $d$-type quarks, respectively.
Since $\sin^2\theta_W \approx 1/4$, the $g_1^{\gamma Z}$ structure function is approximately given by 
    $g_1^{\gamma Z}(x,Q^2)
    \approx \frac19 \sum_q \Delta q^+(x,Q^2)\equiv \frac19 \Delta\Sigma(x,Q^2)$.
With sufficient precision, the combination of $A_{\rm PV}$ and $A_{||}$ for the proton and neutron could enable an extraction of $\Delta u^+$, $\Delta d^+$, and $\Delta s^+$ separately.

\section{Constraints from Polarized DIS Data \label{s:DIS}}

For our baseline analysis, we fit the existing world polarized DIS data on the longitudinal double-spin asymmetries for proton, deuteron, and $^3$He targets.
We restrict the data to the kinematics relevant for this study:~$x < 0.1$ with $Q^2 > m_c^2 \approx 1.69$~GeV$^2$, and, to avoid the nucleon resonance region, $s > 4$~GeV$^2$, where $s$ is the invariant mass squared of the final state hadrons.
The data sets included are from the SLAC~\cite{Anthony:1996mw, Abe:1997cx, Abe:1998wq, Anthony:1999rm, Anthony:2000fn}, EMC~\cite{Ashman:1989ig}, SMC~\cite{Adeva:1998vv, Adeva:1999pa}, COMPASS~\cite{Alekseev:2010hc, Adolph:2015saz,Adolph:2016myg}, and HERMES~\cite{Ackerstaff:1997ws, Airapetian:2007mh} experiments, giving a total number of points $N_{\rm pts}=122$ that survive the cuts.
Note that the variable $y_0=\sqrt{\as N_c/2\pi}\ln(1/x_0)$ that enters the evolution equations (\ref{e:HelEv3_f}) has been fixed using $x_0=0.1$, consistent with the $x$ cut on the data.

\begin{figure}[t]
\centering
\includegraphics[width=0.483\textwidth]{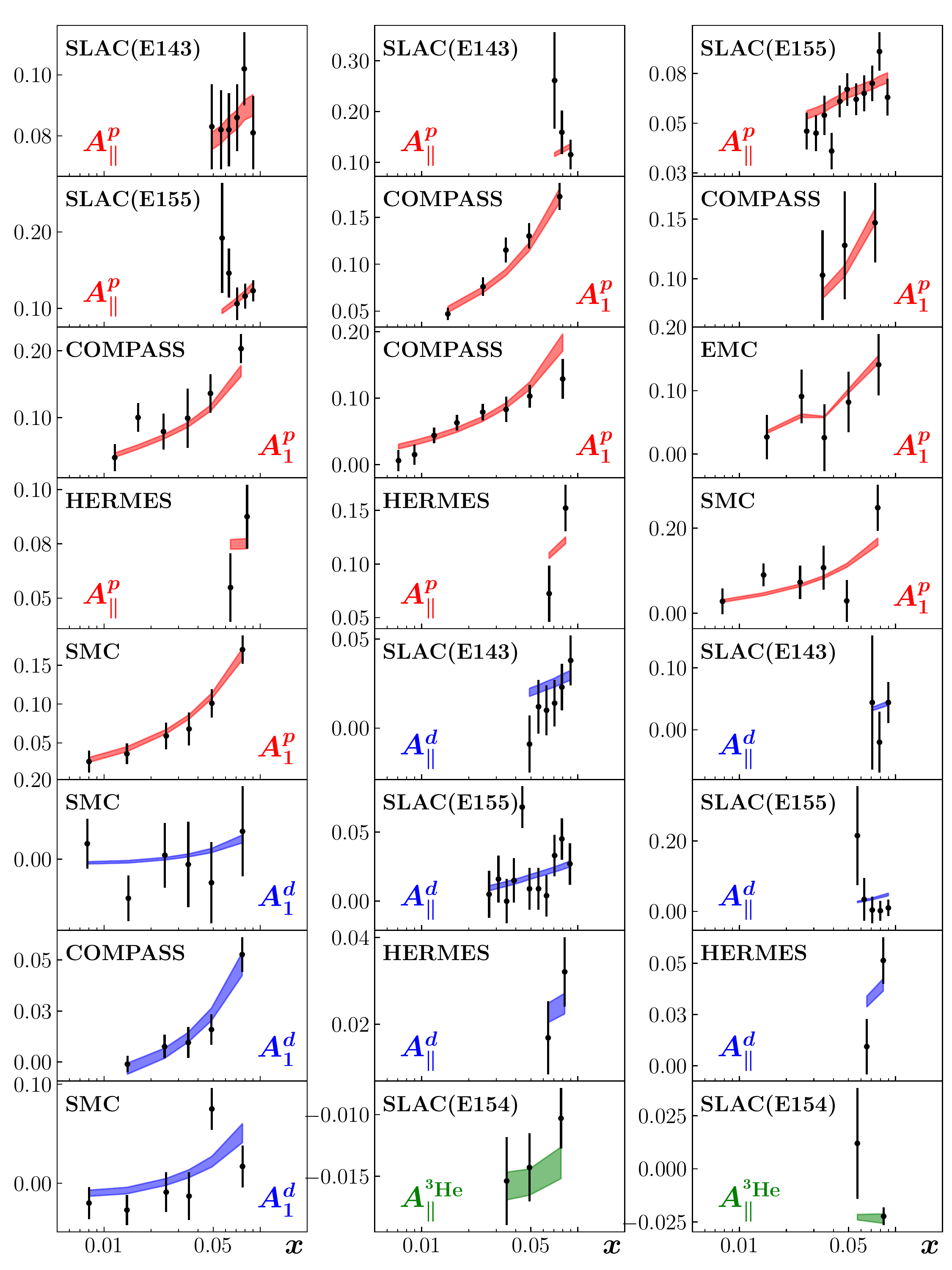}
\caption{Comparison of $A_{||}$ and $A_1$
data (black) at $x < 0.1$ and $Q^2 \in [1.73, 19.70]~$GeV$^2$ with the JAMsmallx fit: proton (red), deuteron (blue), and $^3$He (green).}
\label{f:A1}
\end{figure}

\begin{figure}[t]
\centering
\includegraphics[width=0.483\textwidth]{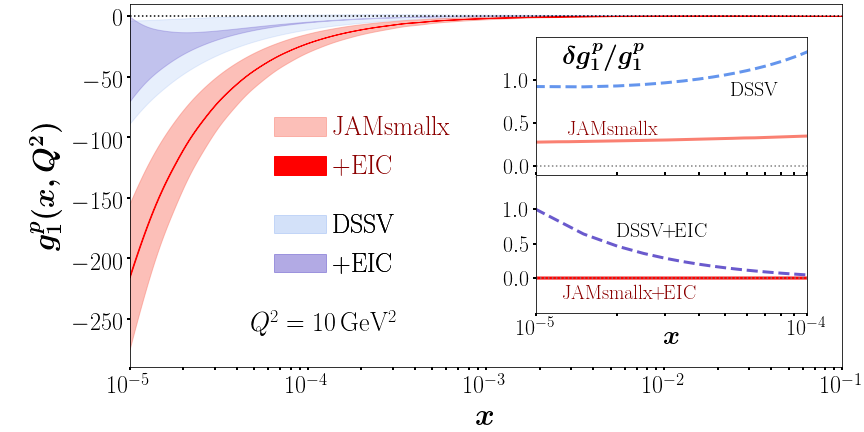}
\caption{
\mbox{JAMsmallx} result for the $g_1^p$ structure function obtained from existing polarized DIS data (light red band) as well as with EIC pseudodata (dark red band). For comparison, we include $g_1^p$ from the DSSV fit to existing data~\cite{deFlorian:2014yva, deFlorian:2019zkl} (light blue band) and with EIC pseudodata at $\sqrt{S}=45$ and 141~GeV~\cite{Aschenauer:2020pdk} (light purple band). The inset gives the relative uncertainty $\delta g_1^p/g_1^p$ for each fit at small $x$.}
\label{f:g1}
\end{figure}

As discussed above, these data alone are not sufficient to extract the individual PDFs $\Delta u^+$, $\Delta d^+$, and $\Delta s^+$.
Instead, we can only constrain the linear combinations of $a_q$, $b_q$, and $c_q$ from Eq.~(\ref{G01}) that enter into the proton $g_1^p$ and neutron $g_1^n$ structure functions (\ref{e:g1LO}).  
This gives effectively six free parameters (in addition to $x_0$ and $\Lambda$). 
That is, the initial conditions for the polarized dipole amplitudes associated with $g_1^p$ and $g_1^n$, respectively, read,
\begin{subequations} \label{e:G0g1}
\begin{align}
G_{p}^{(0)}(s_{10}, \eta) &= a_{p} \, \eta + b_{p} \, s_{10} + c_{p}\,, \\[0.3cm]
     G_{n}^{(0)}(s_{10}, \eta) &= a_{n} \, \eta + b_{n} \, s_{10} + c_{n}\,.
\end{align}%
\end{subequations} 
We determine these parameters using Bayesian inference within the JAM Monte Carlo framework~\cite{Sato:2019yez, Moffat:2021dji} and find the following values: 
$a_{p}=-1.33 \pm 0.30, b_{p}=0.49\pm 0.44, c_{p}=2.24\pm 0.16$, and $a_{n}=-2.47 \pm 0.65, b_{n}=3.03\pm 1.01, c_{n}=0.30\pm 0.36$.
The comparison between our fit (which we refer to as \mbox{``JAMsmallx''}) at 1$\sigma$ confidence level and the $x<0.1$ data on the proton, deuteron, and $^3$He double-spin asymmetries is shown in Fig.~\ref{f:A1}, with the associated $g_1^p$ structure function displayed in Fig.~\ref{f:g1}.
We find a very good fit to the data, with $\chi^2/N_{\rm pts}=1.01$.

\begin{figure}[h]
\centering
\includegraphics[width=0.45\textwidth]{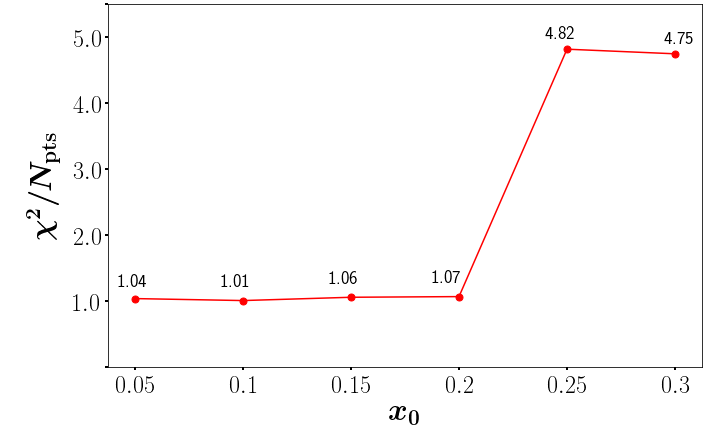}
\caption{Plot of $\chi^2/N_{\rm pts}$ versus $x_0$: the numbers next to the red points indicate the specific $\chi^2/N_{\rm pts}$ values at the given~$x_0$.}
\label{f:chi2}
\end{figure}

The precise value of $x_0$ at which KPS evolution sets in, corresponding to the cut $x<x_0$ applied to the data, is not known {\it a priori}. In Fig.~\ref{f:chi2}, we show $\chi^2/N_{\rm pts}$ for $x_0=\{0.05,0.1,0.15,0.2,0.25,0.3\}$, where $N_{\rm pts}=\{62, 122, 187, 229, 342, 508\}$, respectively. 
We note that for $x_0=0.2$, a few data points from SLAC~E80/E130~\cite{Baum:1983ha} survive the $x<x_0$ cut, and for $x_0=0.25$, also data from Jefferson Lab~\cite{Prok:2014ltt, Parno:2014xzb, Guler:2015hsw, Fersch:2017qrq} survive that $x<x_0$ cut. However, the latter data points are not the sole reason for the increase in $\chi^2/N_{\rm pts}$ when $x_0\ge 0.25$. 
Certain data sets from COMPASS, HERMES, and SLAC that the fit describes well when only $x< 0.2$ points are included also have their individual $\chi^2/N_{\rm pts}$ deteriorate once additional data with $x\ge 0.25$ enter the fit.

The fact that we find good fits up to $x_0=0.2$ introduces an additional systematic uncertainty into the behavior of $g_1^p$ down to $x=10^{-5}$ in Fig.~\ref{f:g1}. The error band in the plot only reflects the uncertainty from the experimental data and not this systematic uncertainty due to the choice of $x_0$.
This ambiguity in $x_0$ indicates that current polarized DIS data have not been measured at small enough $x$ to identify the onset of small-$x$ helicity evolution.
The data do, however, constrain the value of $x_0$ by imposing an upper bound.
Our fit is not expected to work at larger values of $x_0$, where the small-$x$ formalism should become inapplicable.
We find that the data can indeed discriminate this breakdown, with the fit quality $\chi^2/N_{\rm pts}$ degrading substantially for $x_0 \sim 0.25$ due to the inability of the small-$x$ formalism to capture the steep $(1-x)^{n}$ ($n \approx 3$) large-$x$ falloff in the data. We note that the unpolarized evolution resummation parameter $\as\, \ln (1/x)$ at $x=0.01$ is approximately equal to the polarized evolution parameter $\as \, \ln^2 (1/x)$ at $x=0.1$, suggesting comparable accuracy for our helicity evolution with $x_0 = 0.1$ and the unpolarized small-$x$ evolution~\cite{Kuraev:1977fs, Balitsky:1978ic, Balitsky:1995ub, Balitsky:1998ya, Kovchegov:1999yj, Kovchegov:1999ua, JalilianMarian:1997dw, JalilianMarian:1997gr, Weigert:2000gi, Iancu:2001ad, Iancu:2000hn, Ferreiro:2001qy} with the commonly used value of $x_0=0.01$~\cite{Albacete:2010sy, Albacete:2009fh, Kharzeev:2003wz, Albacete:2003iq, Kharzeev:2004yx, Albacete:2004gw}.

We also comment that there exist other quantities with the leading small-$x$ contribution being double-logarithmic in $x$. 
An example would be the flavor non-singlet unpolarized PDFs, which, at small-$x$, are dominated by the QCD Reggeon exchange \cite{Kirschner:1983di, Kirschner:1985cb, Kirschner:1994vc, Kirschner:1994rq, Griffiths:1999dj, Ermolaev:1995fx, Bartels:2003dj}, whose intercept, when evaluated in the DLA for $\as \approx 0.3$, is very close to the phenomenological value of $\alpha_R \approx 0.5$ \cite{Donnachie:1992ny}, as shown in \cite{Itakura:2003jp}.
Moreover, the Reggeon contribution to baryon stopping in heavy ion collisions was also explored in \cite{Itakura:2003jp} (see Fig.~9 there and the discussion around it). 
Surprisingly, no higher-order corrections to the DLA were needed in \cite{Itakura:2003jp} in order to obtain a good agreement with the data.
Therefore, it is possible that the KPS evolution, which is also double-logarithmic at leading order, may give an accurate prediction for the small-$x$ $g_1$ structure function already at DLA, as employed in this work.

A unique feature of our analysis is that KPS evolution {\it predicts} the small-$x$ behavior of helicity PDFs.
This is in contrast to DGLAP evolution, where the $x$ dependence of the PDFs follows from {\it ad hoc} parametrizations at an input scale $Q_0$, with the behavior at small $x$ obtained by extrapolation.
This distinction allows better controlled uncertainties in KPS evolution at small $x$, as Fig.~\ref{f:g1} confirms.  
For the fits to existing data, the relative error $\delta g_1^p/g_1^p$ at small $x$ is $\sim 25\%$ for JAMsmallx and $\sim 100\%$ for the DSSV fit with standard $Q^2$ evolution~\cite{deFlorian:2014yva, deFlorian:2019zkl}.

\section{Impact from EIC Data}
To estimate the impact of future EIC data on the $g_1$ structure function,
we generate pseudodata for $A_{||}$ and $A_{\rm PV}$ for proton, deuteron, and $^3$He beams.  
The fit described in Sec.~\ref{s:DIS} only constrains $g_1^p$ and $g_1^n$, whereas to generate pseudodata simultaneously for $A_{||}$ and $A_{\rm PV}$, one needs $\Delta u^+$, $\Delta d^+$, and $\Delta s^+$ individually.  
Therefore, we set $\Delta s^+=0$ and use isospin symmetry to invert Eq.~(\ref{e:g1LO}) to determine the initial conditions for $\Delta u^+$ and $\Delta d^+$ from those we already extracted for $g_1^p$ and $g_1^n$, such that
\begin{subequations}
\label{e:G0flav}
\begin{align} 
G_u^{(0)}(s_{10},\eta) &= \frac{6}{5}\!\left(4G_{p}^{(0)}(s_{10},\eta)-G_{n}^{(0)}(s_{10},\eta)\right),\\[0.3cm]
G_d^{(0)}(s_{10},\eta) &= \frac{6}{5}\!\left(4G_{n}^{(0)}(s_{10},\eta)-G_{p}^{(0)}(s_{10},\eta)\right),\\[0.3cm]
G_s^{(0)}(s_{10},\eta) &=0\,,
\end{align}
\end{subequations}
with $G_{p}^{(0)}$ and $G_{n}^{(0)}$ 
taken from Eqs.~(\ref{e:G0g1}) for the fit in Sec.~\ref{s:DIS}.
We use the initial conditions (\ref{e:G0flav}) to solve the evolution equations (\ref{e:HelEv3_f}) for the polarized dipole amplitudes corresponding to individual flavors.
Using Eq.~(\ref{Dq+3}), we obtain helicity PDFs which allow us to generate the central values of the EIC pseudodata for $A_{||}$ and $A_{\rm PV}$.  
For the proton, the pseudodata cover center-of-mass~energies $\sqrt{S}= \{ 29, 45, 63, 141 \}$~GeV with integrated luminosity of 100~fb$^{-1}$, while for the deuteron and $^3$He beams the pseudodata span $\sqrt{S}= \{ 29, 66, 89 \}$~GeV with 10~fb$^{-1}$ integrated luminosity. 
These are consistent with the EIC detector design of the Yellow Report, including $2\%$ point-to-point uncorrelated systematic uncertainties~\cite{AbdulKhalek:2021gbh}.
After imposing the kinematic cuts discussed above, 487 data points survive for each of $A_{||}$ and $A_{\rm PV}$, along with the 122 data points from existing polarized DIS data, for a total of 1096 points used in this analysis.

\begin{figure}
\centering
 \includegraphics[width=0.425\textwidth]{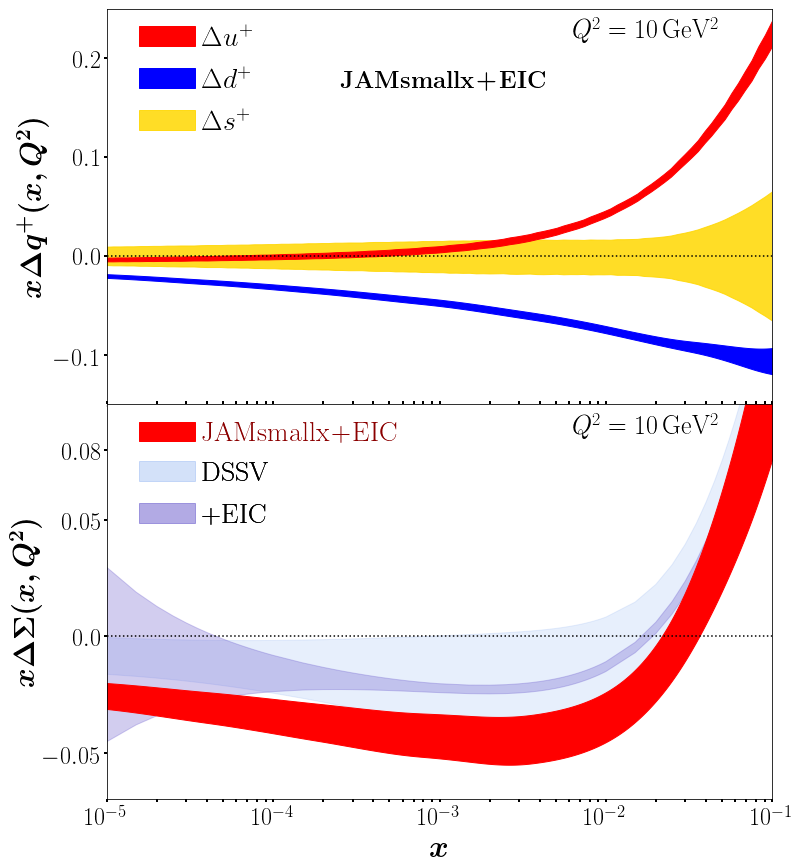}
\caption{{\bf (Top)} Fitted helicity PDFs $x\Delta q^+(x,Q^2)$ from the current JAMsmallx fit to existing polarized DIS data and EIC pseudodata for $A_{||}$ and $A_{\rm PV}$ at $x<0.1$. {\bf (Bottom)} The result for $x\Delta\Sigma(x,Q^2)$ from the same fit (red), compared with that from the DSSV analysis with~\cite{Aschenauer:2020pdk} (light purple) and without~\cite{deFlorian:2014yva, deFlorian:2019zkl} (light blue) the EIC pseudodata.}
\label{f:hPDFs_Sigma}
\end{figure}

We now fit these pseudodata without making any assumptions on the helicity PDFs; in particular, we do not assume $\Delta s^+=0$ in the fit. The inclusion of $A_{\rm PV}$ allows us to extract the individual PDFs $\Delta u^+$, $\Delta d^+$, and $\Delta s^+$ using nine parameters ($a_q$, $b_q$, and $c_q$~({\it cf.}~Eq.~(\ref{G01}))~for each quark flavor) in addition to our choices for $x_0$ and~$\Lambda$.

The results for the extracted helicity PDFs, as well as for the flavor singlet sum $\Delta\Sigma(x,Q^2)$, are shown in Fig.~\ref{f:hPDFs_Sigma}, and $g_1^p$ is given by the dark red band in Fig.~\ref{f:g1}.
Clearly, the EIC pseudodata have a significant impact, reducing the relative uncertainty of $g_1^p$ to the sub-percent level.  
This precision will also allow for a more accurate determination of the starting point $x_0$ of KPS evolution. 
The improved control over the small-$x$ behavior with KPS evolution of the $g_1$ structure function and the helicity PDFs is evident in Figs.~\ref{f:g1} and~\ref{f:hPDFs_Sigma} when compared with the DSSV analysis~\cite{deFlorian:2014yva, deFlorian:2019zkl}, which uses standard DGLAP evolution.
Even after including EIC pseudodata, the relative error of the DSSV+EIC fit~\cite{Aschenauer:2020pdk} for $g_1^p$ grows to $\sim 100 \%$ when one enters the unmeasured region ($x \lesssim 10^{-4}$).
The same trend occurs for $x\Delta\Sigma(x,Q^2)$:~the magnitude of the JAMsmallx+EIC uncertainty band stays relatively constant, while the DSSV+EIC error increases significantly at $x \lesssim 10^{-4}$. 
We emphasize that this is a consequence of DGLAP evolution not being able to prescribe the small-$x$ behavior of PDFs, whereas KPS evolution enables a genuine prediction at small $x$.

\section{Outlook}
In this work, we have demonstrated for the first time that double-spin asymmetries in polarized DIS at $x < 0.1$ can be successfully described using the KPS small-$x$ evolution equations. 
In the future, several extensions can be pursued, such as including $\as \, \ln(1/x)$ corrections to the DLA \cite{Kovchegov:2021lvz} and going beyond the large-$N_c$ limit employed here. 
The former will introduce saturation effects and may permit an extraction of $\Delta G$, while the latter may be studied either in the large-$N_c\, \&\, N_f$ limit~\cite{Kovchegov:2015pbl, Kovchegov:2018znm, Kovchegov:2020hgb} or by using functional methods~\cite{Cougoulic:2019aja}.   
Our formalism can also be extended to SIDIS and $pp$ collisions in order to provide a more universal small-$x$ helicity phenomenology.
The approach we have pioneered here will allow us to achieve well-controlled uncertainties as one extends into the unmeasured small-$x$ region (beyond what even the EIC can reach), a feature that ultimately will be crucial to understanding
the partonic origin of the proton spin.

\section*{Acknowledgments}
This work has been supported by the U.S. Department of Energy, Office of Science, Office of Nuclear Physics under Award Number DE-SC0004286 (DA and YK), No.~DE-AC05-06OR23177 (WM and NS) under which Jefferson Science Associates, LLC, manages and operates Jefferson Lab, the National Science Foundation under Grant No.~PHY-2011763 (DP), and within the framework of the TMD Topical Collaboration. 
The work of NS was supported by the DOE, Office of Science, Office of Nuclear Physics in the Early Career Program.
DA and DP would like to thank C.~Cocuzza and Y.~Zhou for their tutorial on the JAM analysis code. 
We would also like to thank I.~Borsa for providing the results of the analysis in Ref.~\cite{Aschenauer:2020pdk}.
\vspace*{0.9cm}

%


\begin{thebibliography}{83}%
\makeatletter
\providecommand \@ifxundefined [1]{%
 \@ifx{#1\undefined}
}%
\providecommand \@ifnum [1]{%
 \ifnum #1\expandafter \@firstoftwo
 \else \expandafter \@secondoftwo
 \fi
}%
\providecommand \@ifx [1]{%
 \ifx #1\expandafter \@firstoftwo
 \else \expandafter \@secondoftwo
 \fi
}%
\providecommand \natexlab [1]{#1}%
\providecommand \enquote  [1]{``#1''}%
\providecommand \bibnamefont  [1]{#1}%
\providecommand \bibfnamefont [1]{#1}%
\providecommand \citenamefont [1]{#1}%
\providecommand \href@noop [0]{\@secondoftwo}%
\providecommand \href [0]{\begingroup \@sanitize@url \@href}%
\providecommand \@href[1]{\@@startlink{#1}\@@href}%
\providecommand \@@href[1]{\endgroup#1\@@endlink}%
\providecommand \@sanitize@url [0]{\catcode `\\12\catcode `\$12\catcode
  `\&12\catcode `\#12\catcode `\^12\catcode `\_12\catcode `\%12\relax}%
\providecommand \@@startlink[1]{}%
\providecommand \@@endlink[0]{}%
\providecommand \url  [0]{\begingroup\@sanitize@url \@url }%
\providecommand \@url [1]{\endgroup\@href {#1}{\urlprefix }}%
\providecommand \urlprefix  [0]{URL }%
\providecommand \Eprint [0]{\href }%
\providecommand \doibase [0]{http://dx.doi.org/}%
\providecommand \selectlanguage [0]{\@gobble}%
\providecommand \bibinfo  [0]{\@secondoftwo}%
\providecommand \bibfield  [0]{\@secondoftwo}%
\providecommand \translation [1]{[#1]}%
\providecommand \BibitemOpen [0]{}%
\providecommand \bibitemStop [0]{}%
\providecommand \bibitemNoStop [0]{.\EOS\space}%
\providecommand \EOS [0]{\spacefactor3000\relax}%
\providecommand \BibitemShut  [1]{\csname bibitem#1\endcsname}%
\let\auto@bib@innerbib\@empty
\bibitem [{\citenamefont {Jaffe}\ and\ \citenamefont
  {Manohar}(1990)}]{Jaffe:1989jz}%
  \BibitemOpen
  \bibfield  {author} {\bibinfo {author} {\bibfnamefont {R.~L.}\ \bibnamefont
  {Jaffe}}\ and\ \bibinfo {author} {\bibfnamefont {A.}~\bibnamefont
  {Manohar}},\ }\href {\doibase 10.1016/0550-3213(90)90506-9} {\bibfield
  {journal} {\bibinfo  {journal} {Nucl. Phys.}\ }\textbf {\bibinfo {volume}
  {B337}},\ \bibinfo {pages} {509} (\bibinfo {year} {1990})}\BibitemShut
  {NoStop}%
\bibitem [{\citenamefont {Ji}(1997)}]{Ji:1996ek}%
  \BibitemOpen
  \bibfield  {author} {\bibinfo {author} {\bibfnamefont {X.}~\bibnamefont
  {Ji}},\ }\href {\doibase 10.1103/PhysRevLett.78.610} {\bibfield  {journal}
  {\bibinfo  {journal} {Phys. Rev. Lett.}\ }\textbf {\bibinfo {volume} {78}},\
  \bibinfo {pages} {610} (\bibinfo {year} {1997})},\ \Eprint
  {http://arxiv.org/abs/hep-ph/9603249} {arXiv:hep-ph/9603249 [hep-ph]}
  \BibitemShut {NoStop}%
\bibitem [{\citenamefont {Aidala}\ \emph {et~al.}(2013)\citenamefont {Aidala},
  \citenamefont {Bass}, \citenamefont {Hasch},\ and\ \citenamefont
  {Mallot}}]{Aidala:2012mv}%
  \BibitemOpen
  \bibfield  {author} {\bibinfo {author} {\bibfnamefont {C.~A.}\ \bibnamefont
  {Aidala}}, \bibinfo {author} {\bibfnamefont {S.~D.}\ \bibnamefont {Bass}},
  \bibinfo {author} {\bibfnamefont {D.}~\bibnamefont {Hasch}}, \ and\ \bibinfo
  {author} {\bibfnamefont {G.~K.}\ \bibnamefont {Mallot}},\ }\href {\doibase
  10.1103/RevModPhys.85.655} {\bibfield  {journal} {\bibinfo  {journal} {Rev.
  Mod. Phys.}\ }\textbf {\bibinfo {volume} {85}},\ \bibinfo {pages} {655}
  (\bibinfo {year} {2013})},\ \Eprint {http://arxiv.org/abs/1209.2803}
  {arXiv:1209.2803 [hep-ph]} \BibitemShut {NoStop}%
\bibitem [{\citenamefont {Kovchegov}\ \emph {et~al.}(2016)\citenamefont
  {Kovchegov}, \citenamefont {Pitonyak},\ and\ \citenamefont
  {Sievert}}]{Kovchegov:2015pbl}%
  \BibitemOpen
  \bibfield  {author} {\bibinfo {author} {\bibfnamefont {Y.~V.}\ \bibnamefont
  {Kovchegov}}, \bibinfo {author} {\bibfnamefont {D.}~\bibnamefont {Pitonyak}},
  \ and\ \bibinfo {author} {\bibfnamefont {M.~D.}\ \bibnamefont {Sievert}},\
  }\href {\doibase 10.1007/JHEP01(2016)072} {\bibfield  {journal} {\bibinfo
  {journal} {JHEP}\ }\textbf {\bibinfo {volume} {01}},\ \bibinfo {pages} {072}
  (\bibinfo {year} {2016})},\ \bibinfo {note} {[Erratum: JHEP 10, 148
  (2016)]},\ \Eprint {http://arxiv.org/abs/1511.06737} {arXiv:1511.06737
  [hep-ph]} \BibitemShut {NoStop}%
\bibitem [{\citenamefont {Hatta}\ \emph {et~al.}(2017)\citenamefont {Hatta},
  \citenamefont {Nakagawa}, \citenamefont {Yuan}, \citenamefont {Zhao},\ and\
  \citenamefont {Xiao}}]{Hatta:2016aoc}%
  \BibitemOpen
  \bibfield  {author} {\bibinfo {author} {\bibfnamefont {Y.}~\bibnamefont
  {Hatta}}, \bibinfo {author} {\bibfnamefont {Y.}~\bibnamefont {Nakagawa}},
  \bibinfo {author} {\bibfnamefont {F.}~\bibnamefont {Yuan}}, \bibinfo {author}
  {\bibfnamefont {Y.}~\bibnamefont {Zhao}}, \ and\ \bibinfo {author}
  {\bibfnamefont {B.}~\bibnamefont {Xiao}},\ }\href {\doibase
  10.1103/PhysRevD.95.114032} {\bibfield  {journal} {\bibinfo  {journal} {Phys.
  Rev. D}\ }\textbf {\bibinfo {volume} {95}},\ \bibinfo {pages} {114032}
  (\bibinfo {year} {2017})},\ \Eprint {http://arxiv.org/abs/1612.02445}
  {arXiv:1612.02445 [hep-ph]} \BibitemShut {NoStop}%
\bibitem [{\citenamefont {Kovchegov}\ \emph
  {et~al.}(2017{\natexlab{a}})\citenamefont {Kovchegov}, \citenamefont
  {Pitonyak},\ and\ \citenamefont {Sievert}}]{Kovchegov:2016zex}%
  \BibitemOpen
  \bibfield  {author} {\bibinfo {author} {\bibfnamefont {Y.~V.}\ \bibnamefont
  {Kovchegov}}, \bibinfo {author} {\bibfnamefont {D.}~\bibnamefont {Pitonyak}},
  \ and\ \bibinfo {author} {\bibfnamefont {M.~D.}\ \bibnamefont {Sievert}},\
  }\href {\doibase 10.1103/PhysRevD.95.014033} {\bibfield  {journal} {\bibinfo
  {journal} {Phys. Rev. D}\ }\textbf {\bibinfo {volume} {95}},\ \bibinfo
  {pages} {014033} (\bibinfo {year} {2017}{\natexlab{a}})},\ \Eprint
  {http://arxiv.org/abs/1610.06197} {arXiv:1610.06197 [hep-ph]} \BibitemShut
  {NoStop}%
\bibitem [{\citenamefont {Kovchegov}\ \emph
  {et~al.}(2017{\natexlab{b}})\citenamefont {Kovchegov}, \citenamefont
  {Pitonyak},\ and\ \citenamefont {Sievert}}]{Kovchegov:2016weo}%
  \BibitemOpen
  \bibfield  {author} {\bibinfo {author} {\bibfnamefont {Y.~V.}\ \bibnamefont
  {Kovchegov}}, \bibinfo {author} {\bibfnamefont {D.}~\bibnamefont {Pitonyak}},
  \ and\ \bibinfo {author} {\bibfnamefont {M.~D.}\ \bibnamefont {Sievert}},\
  }\href {\doibase 10.1103/PhysRevLett.118.052001} {\bibfield  {journal}
  {\bibinfo  {journal} {Phys. Rev. Lett.}\ }\textbf {\bibinfo {volume} {118}},\
  \bibinfo {pages} {052001} (\bibinfo {year} {2017}{\natexlab{b}})},\ \Eprint
  {http://arxiv.org/abs/1610.06188} {arXiv:1610.06188 [hep-ph]} \BibitemShut
  {NoStop}%
\bibitem [{\citenamefont {Chirilli}(2019)}]{Chirilli:2018kkw}%
  \BibitemOpen
  \bibfield  {author} {\bibinfo {author} {\bibfnamefont {G.~A.}\ \bibnamefont
  {Chirilli}},\ }\href {\doibase 10.1007/JHEP01(2019)118} {\bibfield  {journal}
  {\bibinfo  {journal} {JHEP}\ }\textbf {\bibinfo {volume} {01}},\ \bibinfo
  {pages} {118} (\bibinfo {year} {2019})},\ \Eprint
  {http://arxiv.org/abs/1807.11435} {arXiv:1807.11435 [hep-ph]} \BibitemShut
  {NoStop}%
\bibitem [{\citenamefont {Kovchegov}(2019)}]{Kovchegov:2019rrz}%
  \BibitemOpen
  \bibfield  {author} {\bibinfo {author} {\bibfnamefont {Y.~V.}\ \bibnamefont
  {Kovchegov}},\ }\href {\doibase 10.1007/JHEP03(2019)174} {\bibfield
  {journal} {\bibinfo  {journal} {JHEP}\ }\textbf {\bibinfo {volume} {03}},\
  \bibinfo {pages} {174} (\bibinfo {year} {2019})},\ \Eprint
  {http://arxiv.org/abs/1901.07453} {arXiv:1901.07453 [hep-ph]} \BibitemShut
  {NoStop}%
\bibitem [{\citenamefont {Boussarie}\ \emph {et~al.}(2019)\citenamefont
  {Boussarie}, \citenamefont {Hatta},\ and\ \citenamefont
  {Yuan}}]{Boussarie:2019icw}%
  \BibitemOpen
  \bibfield  {author} {\bibinfo {author} {\bibfnamefont {R.}~\bibnamefont
  {Boussarie}}, \bibinfo {author} {\bibfnamefont {Y.}~\bibnamefont {Hatta}}, \
  and\ \bibinfo {author} {\bibfnamefont {F.}~\bibnamefont {Yuan}},\ }\href
  {\doibase 10.1016/j.physletb.2019.134817} {\bibfield  {journal} {\bibinfo
  {journal} {Phys. Lett. B}\ }\textbf {\bibinfo {volume} {797}},\ \bibinfo
  {pages} {134817} (\bibinfo {year} {2019})},\ \Eprint
  {http://arxiv.org/abs/1904.02693} {arXiv:1904.02693 [hep-ph]} \BibitemShut
  {NoStop}%
\bibitem [{\citenamefont {Chirilli}(2021)}]{Chirilli:2021lif}%
  \BibitemOpen
  \bibfield  {author} {\bibinfo {author} {\bibfnamefont {G.~A.}\ \bibnamefont
  {Chirilli}},\ }\href@noop {} {\  (\bibinfo {year} {2021})},\ \Eprint
  {http://arxiv.org/abs/2101.12744} {arXiv:2101.12744 [hep-ph]} \BibitemShut
  {NoStop}%
\bibitem [{\citenamefont {Kirschner}\ and\ \citenamefont
  {Lipatov}(1983)}]{Kirschner:1983di}%
  \BibitemOpen
  \bibfield  {author} {\bibinfo {author} {\bibfnamefont {R.}~\bibnamefont
  {Kirschner}}\ and\ \bibinfo {author} {\bibfnamefont {L.}~\bibnamefont
  {Lipatov}},\ }\href {\doibase 10.1016/0550-3213(83)90178-5} {\bibfield
  {journal} {\bibinfo  {journal} {Nucl. Phys.}\ }\textbf {\bibinfo {volume}
  {B213}},\ \bibinfo {pages} {122} (\bibinfo {year} {1983})}\BibitemShut
  {NoStop}%
\bibitem [{\citenamefont {Bartels}\ \emph
  {et~al.}(1996{\natexlab{a}})\citenamefont {Bartels}, \citenamefont
  {Ermolaev},\ and\ \citenamefont {Ryskin}}]{Bartels:1995iu}%
  \BibitemOpen
  \bibfield  {author} {\bibinfo {author} {\bibfnamefont {J.}~\bibnamefont
  {Bartels}}, \bibinfo {author} {\bibfnamefont {B.~I.}\ \bibnamefont
  {Ermolaev}}, \ and\ \bibinfo {author} {\bibfnamefont {M.~G.}\ \bibnamefont
  {Ryskin}},\ }\href@noop {} {\bibfield  {journal} {\bibinfo  {journal} {Z.
  Phys. C}\ }\textbf {\bibinfo {volume} {70}},\ \bibinfo {pages} {273}
  (\bibinfo {year} {1996}{\natexlab{a}})},\ \Eprint
  {http://arxiv.org/abs/hep-ph/9507271} {arXiv:hep-ph/9507271 [hep-ph]}
  \BibitemShut {NoStop}%
\bibitem [{\citenamefont {Bartels}\ \emph
  {et~al.}(1996{\natexlab{b}})\citenamefont {Bartels}, \citenamefont
  {Ermolaev},\ and\ \citenamefont {Ryskin}}]{Bartels:1996wc}%
  \BibitemOpen
  \bibfield  {author} {\bibinfo {author} {\bibfnamefont {J.}~\bibnamefont
  {Bartels}}, \bibinfo {author} {\bibfnamefont {B.}~\bibnamefont {Ermolaev}}, \
  and\ \bibinfo {author} {\bibfnamefont {M.}~\bibnamefont {Ryskin}},\ }\href
  {\doibase 10.1007/s002880050285} {\bibfield  {journal} {\bibinfo  {journal}
  {Z. Phys. C}\ }\textbf {\bibinfo {volume} {72}},\ \bibinfo {pages} {627}
  (\bibinfo {year} {1996}{\natexlab{b}})},\ \Eprint
  {http://arxiv.org/abs/hep-ph/9603204} {arXiv:hep-ph/9603204 [hep-ph]}
  \BibitemShut {NoStop}%
\bibitem [{\citenamefont {Kovchegov}\ \emph
  {et~al.}(2017{\natexlab{c}})\citenamefont {Kovchegov}, \citenamefont
  {Pitonyak},\ and\ \citenamefont {Sievert}}]{Kovchegov:2017jxc}%
  \BibitemOpen
  \bibfield  {author} {\bibinfo {author} {\bibfnamefont {Y.~V.}\ \bibnamefont
  {Kovchegov}}, \bibinfo {author} {\bibfnamefont {D.}~\bibnamefont {Pitonyak}},
  \ and\ \bibinfo {author} {\bibfnamefont {M.~D.}\ \bibnamefont {Sievert}},\
  }\href {\doibase 10.1016/j.physletb.2017.06.032} {\bibfield  {journal}
  {\bibinfo  {journal} {Phys. Lett. B}\ }\textbf {\bibinfo {volume} {772}},\
  \bibinfo {pages} {136} (\bibinfo {year} {2017}{\natexlab{c}})},\ \Eprint
  {http://arxiv.org/abs/1703.05809} {arXiv:1703.05809 [hep-ph]} \BibitemShut
  {NoStop}%
\bibitem [{\citenamefont {Kovchegov}\ \emph
  {et~al.}(2017{\natexlab{d}})\citenamefont {Kovchegov}, \citenamefont
  {Pitonyak},\ and\ \citenamefont {Sievert}}]{Kovchegov:2017lsr}%
  \BibitemOpen
  \bibfield  {author} {\bibinfo {author} {\bibfnamefont {Y.~V.}\ \bibnamefont
  {Kovchegov}}, \bibinfo {author} {\bibfnamefont {D.}~\bibnamefont {Pitonyak}},
  \ and\ \bibinfo {author} {\bibfnamefont {M.~D.}\ \bibnamefont {Sievert}},\
  }\href {\doibase 10.1007/JHEP10(2017)198} {\bibfield  {journal} {\bibinfo
  {journal} {JHEP}\ }\textbf {\bibinfo {volume} {10}},\ \bibinfo {pages} {198}
  (\bibinfo {year} {2017}{\natexlab{d}})},\ \Eprint
  {http://arxiv.org/abs/1706.04236} {arXiv:1706.04236 [nucl-th]} \BibitemShut
  {NoStop}%
\bibitem [{\citenamefont {Kovchegov}\ and\ \citenamefont
  {Sievert}(2019)}]{Kovchegov:2018znm}%
  \BibitemOpen
  \bibfield  {author} {\bibinfo {author} {\bibfnamefont {Y.~V.}\ \bibnamefont
  {Kovchegov}}\ and\ \bibinfo {author} {\bibfnamefont {M.~D.}\ \bibnamefont
  {Sievert}},\ }\href {\doibase 10.1103/PhysRevD.99.054032} {\bibfield
  {journal} {\bibinfo  {journal} {Phys. Rev. D}\ }\textbf {\bibinfo {volume}
  {99}},\ \bibinfo {pages} {054032} (\bibinfo {year} {2019})},\ \Eprint
  {http://arxiv.org/abs/1808.09010} {arXiv:1808.09010 [hep-ph]} \BibitemShut
  {NoStop}%
\bibitem [{\citenamefont {Cougoulic}\ and\ \citenamefont
  {Kovchegov}(2019)}]{Cougoulic:2019aja}%
  \BibitemOpen
  \bibfield  {author} {\bibinfo {author} {\bibfnamefont {F.}~\bibnamefont
  {Cougoulic}}\ and\ \bibinfo {author} {\bibfnamefont {Y.~V.}\ \bibnamefont
  {Kovchegov}},\ }\href {\doibase 10.1103/PhysRevD.100.114020} {\bibfield
  {journal} {\bibinfo  {journal} {Phys. Rev. D}\ }\textbf {\bibinfo {volume}
  {100}},\ \bibinfo {pages} {114020} (\bibinfo {year} {2019})},\ \Eprint
  {http://arxiv.org/abs/1910.04268} {arXiv:1910.04268 [hep-ph]} \BibitemShut
  {NoStop}%
\bibitem [{\citenamefont {Gribov}\ and\ \citenamefont
  {Lipatov}(1972)}]{Gribov:1972ri}%
  \BibitemOpen
  \bibfield  {author} {\bibinfo {author} {\bibfnamefont {V.~N.}\ \bibnamefont
  {Gribov}}\ and\ \bibinfo {author} {\bibfnamefont {L.~N.}\ \bibnamefont
  {Lipatov}},\ }\href@noop {} {\bibfield  {journal} {\bibinfo  {journal} {Sov.
  J. Nucl. Phys.}\ }\textbf {\bibinfo {volume} {15}},\ \bibinfo {pages} {438}
  (\bibinfo {year} {1972})}\BibitemShut {NoStop}%
\bibitem [{\citenamefont {Altarelli}\ and\ \citenamefont
  {Parisi}(1977)}]{Altarelli:1977zs}%
  \BibitemOpen
  \bibfield  {author} {\bibinfo {author} {\bibfnamefont {G.}~\bibnamefont
  {Altarelli}}\ and\ \bibinfo {author} {\bibfnamefont {G.}~\bibnamefont
  {Parisi}},\ }\href {\doibase 10.1016/0550-3213(77)90384-4} {\bibfield
  {journal} {\bibinfo  {journal} {Nucl. Phys.}\ }\textbf {\bibinfo {volume}
  {B126}},\ \bibinfo {pages} {298} (\bibinfo {year} {1977})}\BibitemShut
  {NoStop}%
\bibitem [{\citenamefont {Dokshitzer}(1977)}]{Dokshitzer:1977sg}%
  \BibitemOpen
  \bibfield  {author} {\bibinfo {author} {\bibfnamefont {Y.~L.}\ \bibnamefont
  {Dokshitzer}},\ }\href@noop {} {\bibfield  {journal} {\bibinfo  {journal}
  {Sov. Phys. JETP}\ }\textbf {\bibinfo {volume} {46}},\ \bibinfo {pages} {641}
  (\bibinfo {year} {1977})}\BibitemShut {NoStop}%
\bibitem [{\citenamefont {Balitsky}(1996)}]{Balitsky:1995ub}%
  \BibitemOpen
  \bibfield  {author} {\bibinfo {author} {\bibfnamefont {I.}~\bibnamefont
  {Balitsky}},\ }\href {\doibase 10.1016/0550-3213(95)00638-9} {\bibfield
  {journal} {\bibinfo  {journal} {Nucl. Phys.}\ }\textbf {\bibinfo {volume}
  {B463}},\ \bibinfo {pages} {99} (\bibinfo {year} {1996})},\ \Eprint
  {http://arxiv.org/abs/hep-ph/9509348} {arXiv:hep-ph/9509348 [hep-ph]}
  \BibitemShut {NoStop}%
\bibitem [{\citenamefont {Balitsky}(1999)}]{Balitsky:1998ya}%
  \BibitemOpen
  \bibfield  {author} {\bibinfo {author} {\bibfnamefont {I.}~\bibnamefont
  {Balitsky}},\ }\href {\doibase 10.1103/PhysRevD.60.014020} {\bibfield
  {journal} {\bibinfo  {journal} {Phys. Rev. D}\ }\textbf {\bibinfo {volume}
  {60}},\ \bibinfo {pages} {014020} (\bibinfo {year} {1999})},\ \Eprint
  {http://arxiv.org/abs/hep-ph/9812311} {arXiv:hep-ph/9812311} \BibitemShut
  {NoStop}%
\bibitem [{\citenamefont {Kovchegov}(1999)}]{Kovchegov:1999yj}%
  \BibitemOpen
  \bibfield  {author} {\bibinfo {author} {\bibfnamefont {Y.~V.}\ \bibnamefont
  {Kovchegov}},\ }\href {\doibase 10.1103/PhysRevD.60.034008} {\bibfield
  {journal} {\bibinfo  {journal} {Phys. Rev. D}\ }\textbf {\bibinfo {volume}
  {60}},\ \bibinfo {pages} {034008} (\bibinfo {year} {1999})},\ \Eprint
  {http://arxiv.org/abs/hep-ph/9901281} {arXiv:hep-ph/9901281} \BibitemShut
  {NoStop}%
\bibitem [{\citenamefont {Kovchegov}(2000)}]{Kovchegov:1999ua}%
  \BibitemOpen
  \bibfield  {author} {\bibinfo {author} {\bibfnamefont {Y.~V.}\ \bibnamefont
  {Kovchegov}},\ }\href {\doibase 10.1103/PhysRevD.61.074018} {\bibfield
  {journal} {\bibinfo  {journal} {Phys. Rev. D}\ }\textbf {\bibinfo {volume}
  {61}},\ \bibinfo {pages} {074018} (\bibinfo {year} {2000})},\ \Eprint
  {http://arxiv.org/abs/hep-ph/9905214} {arXiv:hep-ph/9905214} \BibitemShut
  {NoStop}%
\bibitem [{\citenamefont {Kovchegov}\ and\ \citenamefont
  {Weigert}(2007)}]{Kovchegov:2006vj}%
  \BibitemOpen
  \bibfield  {author} {\bibinfo {author} {\bibfnamefont {Y.~V.}\ \bibnamefont
  {Kovchegov}}\ and\ \bibinfo {author} {\bibfnamefont {H.}~\bibnamefont
  {Weigert}},\ }\href {\doibase 10.1016/j.nuclphysa.2006.10.075} {\bibfield
  {journal} {\bibinfo  {journal} {Nucl. Phys.}\ }\textbf {\bibinfo {volume}
  {A784}},\ \bibinfo {pages} {188} (\bibinfo {year} {2007})},\ \Eprint
  {http://arxiv.org/abs/hep-ph/0609090} {arXiv:hep-ph/0609090} \BibitemShut
  {NoStop}%
\bibitem [{\citenamefont {Balitsky}(2007)}]{Balitsky:2006wa}%
  \BibitemOpen
  \bibfield  {author} {\bibinfo {author} {\bibfnamefont {I.}~\bibnamefont
  {Balitsky}},\ }\href {\doibase 10.1103/PhysRevD.75.014001} {\bibfield
  {journal} {\bibinfo  {journal} {Phys. Rev. D}\ }\textbf {\bibinfo {volume}
  {75}},\ \bibinfo {pages} {014001} (\bibinfo {year} {2007})},\ \Eprint
  {http://arxiv.org/abs/hep-ph/0609105} {arXiv:hep-ph/0609105} \BibitemShut
  {NoStop}%
\bibitem [{\citenamefont {Albacete}\ \emph {et~al.}(2011)\citenamefont
  {Albacete}, \citenamefont {Armesto}, \citenamefont {Milhano}, \citenamefont
  {Quiroga-Arias},\ and\ \citenamefont {Salgado}}]{Albacete:2010sy}%
  \BibitemOpen
  \bibfield  {author} {\bibinfo {author} {\bibfnamefont {J.~L.}\ \bibnamefont
  {Albacete}}, \bibinfo {author} {\bibfnamefont {N.}~\bibnamefont {Armesto}},
  \bibinfo {author} {\bibfnamefont {J.~G.}\ \bibnamefont {Milhano}}, \bibinfo
  {author} {\bibfnamefont {P.}~\bibnamefont {Quiroga-Arias}}, \ and\ \bibinfo
  {author} {\bibfnamefont {C.~A.}\ \bibnamefont {Salgado}},\ }\href {\doibase
  10.1140/epjc/s10052-011-1705-3} {\bibfield  {journal} {\bibinfo  {journal}
  {Eur. Phys. J. C}\ }\textbf {\bibinfo {volume} {71}},\ \bibinfo {pages}
  {1705} (\bibinfo {year} {2011})},\ \Eprint {http://arxiv.org/abs/1012.4408}
  {arXiv:1012.4408 [hep-ph]} \BibitemShut {NoStop}%
\bibitem [{\citenamefont {Albacete}\ \emph {et~al.}(2009)\citenamefont
  {Albacete}, \citenamefont {Armesto}, \citenamefont {Milhano},\ and\
  \citenamefont {Salgado}}]{Albacete:2009fh}%
  \BibitemOpen
  \bibfield  {author} {\bibinfo {author} {\bibfnamefont {J.~L.}\ \bibnamefont
  {Albacete}}, \bibinfo {author} {\bibfnamefont {N.}~\bibnamefont {Armesto}},
  \bibinfo {author} {\bibfnamefont {J.~G.}\ \bibnamefont {Milhano}}, \ and\
  \bibinfo {author} {\bibfnamefont {C.~A.}\ \bibnamefont {Salgado}},\ }\href
  {\doibase 10.1103/PhysRevD.80.034031} {\bibfield  {journal} {\bibinfo
  {journal} {Phys. Rev. D}\ }\textbf {\bibinfo {volume} {80}},\ \bibinfo
  {pages} {034031} (\bibinfo {year} {2009})},\ \Eprint
  {http://arxiv.org/abs/0902.1112} {arXiv:0902.1112 [hep-ph]} \BibitemShut
  {NoStop}%
\bibitem [{\citenamefont {Beuf}\ \emph {et~al.}(2020)\citenamefont {Beuf},
  \citenamefont {H\"anninen}, \citenamefont {Lappi},\ and\ \citenamefont
  {M\"antysaari}}]{Beuf:2020dxl}%
  \BibitemOpen
  \bibfield  {author} {\bibinfo {author} {\bibfnamefont {G.}~\bibnamefont
  {Beuf}}, \bibinfo {author} {\bibfnamefont {H.}~\bibnamefont {H\"anninen}},
  \bibinfo {author} {\bibfnamefont {T.}~\bibnamefont {Lappi}}, \ and\ \bibinfo
  {author} {\bibfnamefont {H.}~\bibnamefont {M\"antysaari}},\ }\href {\doibase
  10.1103/PhysRevD.102.074028} {\bibfield  {journal} {\bibinfo  {journal}
  {Phys. Rev. D}\ }\textbf {\bibinfo {volume} {102}},\ \bibinfo {pages}
  {074028} (\bibinfo {year} {2020})},\ \Eprint
  {http://arxiv.org/abs/2007.01645} {arXiv:2007.01645 [hep-ph]} \BibitemShut
  {NoStop}%
\bibitem [{\citenamefont {Bl{\"u}mlein}\ and\ \citenamefont
  {Vogt}(1996)}]{Blumlein:1996hb}%
  \BibitemOpen
  \bibfield  {author} {\bibinfo {author} {\bibfnamefont {J.}~\bibnamefont
  {Bl{\"u}mlein}}\ and\ \bibinfo {author} {\bibfnamefont {A.}~\bibnamefont
  {Vogt}},\ }\href {\doibase 10.1016/0370-2693(96)00958-6} {\bibfield
  {journal} {\bibinfo  {journal} {Phys. Lett. B}\ }\textbf {\bibinfo {volume}
  {386}},\ \bibinfo {pages} {350} (\bibinfo {year} {1996})},\ \Eprint
  {http://arxiv.org/abs/hep-ph/9606254} {arXiv:hep-ph/9606254} \BibitemShut
  {NoStop}%
\bibitem [{\citenamefont {Kovchegov}\ and\ \citenamefont
  {Tawabutr}(2020)}]{Kovchegov:2020hgb}%
  \BibitemOpen
  \bibfield  {author} {\bibinfo {author} {\bibfnamefont {Y.~V.}\ \bibnamefont
  {Kovchegov}}\ and\ \bibinfo {author} {\bibfnamefont {Y.}~\bibnamefont
  {Tawabutr}},\ }\href {\doibase 10.1007/JHEP08(2020)014} {\bibfield  {journal}
  {\bibinfo  {journal} {JHEP}\ }\textbf {\bibinfo {volume} {08}},\ \bibinfo
  {pages} {014} (\bibinfo {year} {2020})},\ \Eprint
  {http://arxiv.org/abs/2005.07285} {arXiv:2005.07285 [hep-ph]} \BibitemShut
  {NoStop}%
\bibitem [{\citenamefont {Cocuzza}\ \emph {et~al.}(2021)\citenamefont
  {Cocuzza}, \citenamefont {Ethier}, \citenamefont {Melnitchouk}, \citenamefont
  {Metz},\ and\ \citenamefont {Sato}}]{Cocuzza21}%
  \BibitemOpen
  \bibfield  {author} {\bibinfo {author} {\bibfnamefont {C.}~\bibnamefont
  {Cocuzza}}, \bibinfo {author} {\bibfnamefont {J.~J.}\ \bibnamefont {Ethier}},
  \bibinfo {author} {\bibfnamefont {W.}~\bibnamefont {Melnitchouk}}, \bibinfo
  {author} {\bibfnamefont {A.}~\bibnamefont {Metz}}, \ and\ \bibinfo {author}
  {\bibfnamefont {N.}~\bibnamefont {Sato}},\ }\href@noop {} {\bibfield
  {journal} {\bibinfo  {journal} {in preparation}\ } (\bibinfo {year}
  {2021})}\BibitemShut {NoStop}%
\bibitem [{\citenamefont {Sato}\ \emph {et~al.}(2020)\citenamefont {Sato},
  \citenamefont {Andres}, \citenamefont {Ethier},\ and\ \citenamefont
  {Melnitchouk}}]{Sato:2019yez}%
  \BibitemOpen
  \bibfield  {author} {\bibinfo {author} {\bibfnamefont {N.}~\bibnamefont
  {Sato}}, \bibinfo {author} {\bibfnamefont {C.}~\bibnamefont {Andres}},
  \bibinfo {author} {\bibfnamefont {J.~J.}\ \bibnamefont {Ethier}}, \ and\
  \bibinfo {author} {\bibfnamefont {W.}~\bibnamefont {Melnitchouk}} (\bibinfo
  {collaboration} {JAM}),\ }\href {\doibase 10.1103/PhysRevD.101.074020}
  {\bibfield  {journal} {\bibinfo  {journal} {Phys. Rev. D}\ }\textbf {\bibinfo
  {volume} {101}},\ \bibinfo {pages} {074020} (\bibinfo {year} {2020})},\
  \Eprint {http://arxiv.org/abs/1905.03788} {arXiv:1905.03788 [hep-ph]}
  \BibitemShut {NoStop}%
\bibitem [{\citenamefont {Jimenez-Delgado}\ \emph {et~al.}(2014)\citenamefont
  {Jimenez-Delgado}, \citenamefont {Accardi},\ and\ \citenamefont
  {Melnitchouk}}]{Jimenez-Delgado:2013boa}%
  \BibitemOpen
  \bibfield  {author} {\bibinfo {author} {\bibfnamefont {P.}~\bibnamefont
  {Jimenez-Delgado}}, \bibinfo {author} {\bibfnamefont {A.}~\bibnamefont
  {Accardi}}, \ and\ \bibinfo {author} {\bibfnamefont {W.}~\bibnamefont
  {Melnitchouk}},\ }\href {\doibase 10.1103/PhysRevD.89.034025} {\bibfield
  {journal} {\bibinfo  {journal} {Phys. Rev. D}\ }\textbf {\bibinfo {volume}
  {89}},\ \bibinfo {pages} {034025} (\bibinfo {year} {2014})},\ \Eprint
  {http://arxiv.org/abs/1310.3734} {arXiv:1310.3734 [hep-ph]} \BibitemShut
  {NoStop}%
\bibitem [{\citenamefont {Sato}\ \emph {et~al.}(2016)\citenamefont {Sato},
  \citenamefont {Melnitchouk}, \citenamefont {Kuhn}, \citenamefont {Ethier},\
  and\ \citenamefont {Accardi}}]{Sato:2016tuz}%
  \BibitemOpen
  \bibfield  {author} {\bibinfo {author} {\bibfnamefont {N.}~\bibnamefont
  {Sato}}, \bibinfo {author} {\bibfnamefont {W.}~\bibnamefont {Melnitchouk}},
  \bibinfo {author} {\bibfnamefont {S.}~\bibnamefont {Kuhn}}, \bibinfo {author}
  {\bibfnamefont {J.}~\bibnamefont {Ethier}}, \ and\ \bibinfo {author}
  {\bibfnamefont {A.}~\bibnamefont {Accardi}},\ }\href {\doibase
  10.1103/PhysRevD.93.074005} {\bibfield  {journal} {\bibinfo  {journal} {Phys.
  Rev. D}\ }\textbf {\bibinfo {volume} {93}},\ \bibinfo {pages} {074005}
  (\bibinfo {year} {2016})},\ \Eprint {http://arxiv.org/abs/1601.07782}
  {arXiv:1601.07782 [hep-ph]} \BibitemShut {NoStop}%
\bibitem [{\citenamefont {Ethier}\ \emph {et~al.}(2017)\citenamefont {Ethier},
  \citenamefont {Sato},\ and\ \citenamefont {Melnitchouk}}]{Ethier:2017zbq}%
  \BibitemOpen
  \bibfield  {author} {\bibinfo {author} {\bibfnamefont {J.~J.}\ \bibnamefont
  {Ethier}}, \bibinfo {author} {\bibfnamefont {N.}~\bibnamefont {Sato}}, \ and\
  \bibinfo {author} {\bibfnamefont {W.}~\bibnamefont {Melnitchouk}},\ }\href
  {\doibase 10.1103/PhysRevLett.119.132001} {\bibfield  {journal} {\bibinfo
  {journal} {Phys. Rev. Lett.}\ }\textbf {\bibinfo {volume} {119}},\ \bibinfo
  {pages} {132001} (\bibinfo {year} {2017})},\ \Eprint
  {http://arxiv.org/abs/1705.05889} {arXiv:1705.05889 [hep-ph]} \BibitemShut
  {NoStop}%
\bibitem [{\citenamefont {Moffat}\ \emph {et~al.}(2021)\citenamefont {Moffat},
  \citenamefont {Melnitchouk}, \citenamefont {Rogers},\ and\ \citenamefont
  {Sato}}]{Moffat:2021dji}%
  \BibitemOpen
  \bibfield  {author} {\bibinfo {author} {\bibfnamefont {E.}~\bibnamefont
  {Moffat}}, \bibinfo {author} {\bibfnamefont {W.}~\bibnamefont {Melnitchouk}},
  \bibinfo {author} {\bibfnamefont {T.}~\bibnamefont {Rogers}}, \ and\ \bibinfo
  {author} {\bibfnamefont {N.}~\bibnamefont {Sato}},\ }\href@noop {} {\
  (\bibinfo {year} {2021})},\ \Eprint {http://arxiv.org/abs/2101.04664}
  {arXiv:2101.04664 [hep-ph]} \BibitemShut {NoStop}%
\bibitem [{\citenamefont {Hobbs}\ and\ \citenamefont
  {Melnitchouk}(2008)}]{Hobbs:2008mm}%
  \BibitemOpen
  \bibfield  {author} {\bibinfo {author} {\bibfnamefont {T.}~\bibnamefont
  {Hobbs}}\ and\ \bibinfo {author} {\bibfnamefont {W.}~\bibnamefont
  {Melnitchouk}},\ }\href {\doibase 10.1103/PhysRevD.77.114023} {\bibfield
  {journal} {\bibinfo  {journal} {Phys. Rev. D}\ }\textbf {\bibinfo {volume}
  {77}},\ \bibinfo {pages} {114023} (\bibinfo {year} {2008})},\ \Eprint
  {http://arxiv.org/abs/0801.4791} {arXiv:0801.4791 [hep-ph]} \BibitemShut
  {NoStop}%
\bibitem [{\citenamefont {Zhao}\ \emph {et~al.}(2017)\citenamefont {Zhao},
  \citenamefont {Deshpande}, \citenamefont {Huang}, \citenamefont {Kumar},\
  and\ \citenamefont {Riordan}}]{Zhao:2016rfu}%
  \BibitemOpen
  \bibfield  {author} {\bibinfo {author} {\bibfnamefont {Y.}~\bibnamefont
  {Zhao}}, \bibinfo {author} {\bibfnamefont {A.}~\bibnamefont {Deshpande}},
  \bibinfo {author} {\bibfnamefont {J.}~\bibnamefont {Huang}}, \bibinfo
  {author} {\bibfnamefont {K.}~\bibnamefont {Kumar}}, \ and\ \bibinfo {author}
  {\bibfnamefont {S.}~\bibnamefont {Riordan}},\ }\href {\doibase
  10.1140/epja/i2017-12245-2} {\bibfield  {journal} {\bibinfo  {journal} {Eur.
  Phys. J. A}\ }\textbf {\bibinfo {volume} {53}},\ \bibinfo {pages} {55}
  (\bibinfo {year} {2017})},\ \Eprint {http://arxiv.org/abs/1612.06927}
  {arXiv:1612.06927 [nucl-ex]} \BibitemShut {NoStop}%
\bibitem [{\citenamefont {Anthony}\ \emph {et~al.}(1996)\citenamefont {Anthony}
  \emph {et~al.}}]{Anthony:1996mw}%
  \BibitemOpen
  \bibfield  {author} {\bibinfo {author} {\bibfnamefont {P.~L.}\ \bibnamefont
  {Anthony}} \emph {et~al.} (\bibinfo {collaboration} {E142 Collaboration}),\
  }\href {\doibase 10.1103/PhysRevD.54.6620} {\bibfield  {journal} {\bibinfo
  {journal} {Phys. Rev. D}\ }\textbf {\bibinfo {volume} {54}},\ \bibinfo
  {pages} {6620} (\bibinfo {year} {1996})},\ \Eprint
  {http://arxiv.org/abs/hep-ex/9610007} {arXiv:hep-ex/9610007} \BibitemShut
  {NoStop}%
\bibitem [{\citenamefont {Abe}\ \emph {et~al.}(1997)\citenamefont {Abe} \emph
  {et~al.}}]{Abe:1997cx}%
  \BibitemOpen
  \bibfield  {author} {\bibinfo {author} {\bibfnamefont {K.}~\bibnamefont
  {Abe}} \emph {et~al.} (\bibinfo {collaboration} {E154 Collaboration}),\
  }\href {\doibase 10.1103/PhysRevLett.79.26} {\bibfield  {journal} {\bibinfo
  {journal} {Phys. Rev. Lett.}\ }\textbf {\bibinfo {volume} {79}},\ \bibinfo
  {pages} {26} (\bibinfo {year} {1997})},\ \Eprint
  {http://arxiv.org/abs/hep-ex/9705012} {arXiv:hep-ex/9705012} \BibitemShut
  {NoStop}%
\bibitem [{\citenamefont {Abe}\ \emph {et~al.}(1998)\citenamefont {Abe} \emph
  {et~al.}}]{Abe:1998wq}%
  \BibitemOpen
  \bibfield  {author} {\bibinfo {author} {\bibfnamefont {K.}~\bibnamefont
  {Abe}} \emph {et~al.} (\bibinfo {collaboration} {E143 Collaboration}),\
  }\href {\doibase 10.1103/PhysRevD.58.112003} {\bibfield  {journal} {\bibinfo
  {journal} {Phys. Rev. D}\ }\textbf {\bibinfo {volume} {58}},\ \bibinfo
  {pages} {112003} (\bibinfo {year} {1998})},\ \Eprint
  {http://arxiv.org/abs/hep-ph/9802357} {arXiv:hep-ph/9802357} \BibitemShut
  {NoStop}%
\bibitem [{\citenamefont {Anthony}\ \emph {et~al.}(1999)\citenamefont {Anthony}
  \emph {et~al.}}]{Anthony:1999rm}%
  \BibitemOpen
  \bibfield  {author} {\bibinfo {author} {\bibfnamefont {P.~L.}\ \bibnamefont
  {Anthony}} \emph {et~al.} (\bibinfo {collaboration} {E155 Collaboration}),\
  }\href {\doibase 10.1016/S0370-2693(99)00940-5} {\bibfield  {journal}
  {\bibinfo  {journal} {Phys. Lett. B}\ }\textbf {\bibinfo {volume} {463}},\
  \bibinfo {pages} {339} (\bibinfo {year} {1999})},\ \Eprint
  {http://arxiv.org/abs/hep-ex/9904002} {arXiv:hep-ex/9904002} \BibitemShut
  {NoStop}%
\bibitem [{\citenamefont {Anthony}\ \emph {et~al.}(2000)\citenamefont {Anthony}
  \emph {et~al.}}]{Anthony:2000fn}%
  \BibitemOpen
  \bibfield  {author} {\bibinfo {author} {\bibfnamefont {P.~L.}\ \bibnamefont
  {Anthony}} \emph {et~al.} (\bibinfo {collaboration} {E155 Collaboration}),\
  }\href {\doibase 10.1016/S0370-2693(00)01014-5} {\bibfield  {journal}
  {\bibinfo  {journal} {Phys. Lett. B}\ }\textbf {\bibinfo {volume} {493}},\
  \bibinfo {pages} {19} (\bibinfo {year} {2000})},\ \Eprint
  {http://arxiv.org/abs/hep-ph/0007248} {arXiv:hep-ph/0007248} \BibitemShut
  {NoStop}%
\bibitem [{\citenamefont {Ashman}\ \emph {et~al.}(1989)\citenamefont {Ashman}
  \emph {et~al.}}]{Ashman:1989ig}%
  \BibitemOpen
  \bibfield  {author} {\bibinfo {author} {\bibfnamefont {J.}~\bibnamefont
  {Ashman}} \emph {et~al.} (\bibinfo {collaboration} {European Muon
  Collaboration}),\ }\href {\doibase 10.1016/0550-3213(89)90089-8} {\bibfield
  {journal} {\bibinfo  {journal} {Nucl. Phys.}\ }\textbf {\bibinfo {volume}
  {B328}},\ \bibinfo {pages} {1} (\bibinfo {year} {1989})}\BibitemShut
  {NoStop}%
\bibitem [{\citenamefont {Adeva}\ \emph {et~al.}(1998)\citenamefont {Adeva}
  \emph {et~al.}}]{Adeva:1998vv}%
  \BibitemOpen
  \bibfield  {author} {\bibinfo {author} {\bibfnamefont {B.}~\bibnamefont
  {Adeva}} \emph {et~al.} (\bibinfo {collaboration} {Spin Muon
  Collaboration}),\ }\href {\doibase 10.1103/PhysRevD.58.112001} {\bibfield
  {journal} {\bibinfo  {journal} {Phys. Rev. D}\ }\textbf {\bibinfo {volume}
  {58}},\ \bibinfo {pages} {112001} (\bibinfo {year} {1998})}\BibitemShut
  {NoStop}%
\bibitem [{\citenamefont {Adeva}\ \emph {et~al.}(1999)\citenamefont {Adeva}
  \emph {et~al.}}]{Adeva:1999pa}%
  \BibitemOpen
  \bibfield  {author} {\bibinfo {author} {\bibfnamefont {B.}~\bibnamefont
  {Adeva}} \emph {et~al.} (\bibinfo {collaboration} {Spin Muon
  Collaboration}),\ }\href {\doibase 10.1103/PhysRevD.60.072004} {\bibfield
  {journal} {\bibinfo  {journal} {Phys. Rev. D}\ }\textbf {\bibinfo {volume}
  {60}},\ \bibinfo {pages} {072004} (\bibinfo {year} {1999})},\ \bibinfo {note}
  {[Erratum: Phys.Rev.D 62, 079902 (2000)]}\BibitemShut {NoStop}%
\bibitem [{\citenamefont {Alekseev}\ \emph {et~al.}(2010)\citenamefont
  {Alekseev} \emph {et~al.}}]{Alekseev:2010hc}%
  \BibitemOpen
  \bibfield  {author} {\bibinfo {author} {\bibfnamefont {M.~G.}\ \bibnamefont
  {Alekseev}} \emph {et~al.} (\bibinfo {collaboration} {COMPASS
  Collaboration}),\ }\href {\doibase 10.1016/j.physletb.2010.05.069} {\bibfield
   {journal} {\bibinfo  {journal} {Phys. Lett. B}\ }\textbf {\bibinfo {volume}
  {690}},\ \bibinfo {pages} {466} (\bibinfo {year} {2010})},\ \Eprint
  {http://arxiv.org/abs/1001.4654} {arXiv:1001.4654 [hep-ex]} \BibitemShut
  {NoStop}%
\bibitem [{\citenamefont {Adolph}\ \emph {et~al.}(2016)\citenamefont {Adolph}
  \emph {et~al.}}]{Adolph:2015saz}%
  \BibitemOpen
  \bibfield  {author} {\bibinfo {author} {\bibfnamefont {C.}~\bibnamefont
  {Adolph}} \emph {et~al.} (\bibinfo {collaboration} {COMPASS Collaboration}),\
  }\href {\doibase 10.1016/j.physletb.2015.11.064} {\bibfield  {journal}
  {\bibinfo  {journal} {Phys. Lett. B}\ }\textbf {\bibinfo {volume} {753}},\
  \bibinfo {pages} {18} (\bibinfo {year} {2016})},\ \Eprint
  {http://arxiv.org/abs/1503.08935} {arXiv:1503.08935 [hep-ex]} \BibitemShut
  {NoStop}%
\bibitem [{\citenamefont {Adolph}\ \emph {et~al.}(2017)\citenamefont {Adolph}
  \emph {et~al.}}]{Adolph:2016myg}%
  \BibitemOpen
  \bibfield  {author} {\bibinfo {author} {\bibfnamefont {C.}~\bibnamefont
  {Adolph}} \emph {et~al.} (\bibinfo {collaboration} {COMPASS}),\ }\href
  {\doibase 10.1016/j.physletb.2017.03.018} {\bibfield  {journal} {\bibinfo
  {journal} {Phys. Lett. B}\ }\textbf {\bibinfo {volume} {769}},\ \bibinfo
  {pages} {34} (\bibinfo {year} {2017})},\ \Eprint
  {http://arxiv.org/abs/1612.00620} {arXiv:1612.00620 [hep-ex]} \BibitemShut
  {NoStop}%
\bibitem [{\citenamefont {Ackerstaff}\ \emph {et~al.}(1997)\citenamefont
  {Ackerstaff} \emph {et~al.}}]{Ackerstaff:1997ws}%
  \BibitemOpen
  \bibfield  {author} {\bibinfo {author} {\bibfnamefont {K.}~\bibnamefont
  {Ackerstaff}} \emph {et~al.} (\bibinfo {collaboration} {HERMES
  Collaboration}),\ }\href {\doibase 10.1016/S0370-2693(97)00611-4} {\bibfield
  {journal} {\bibinfo  {journal} {Phys. Lett. B}\ }\textbf {\bibinfo {volume}
  {404}},\ \bibinfo {pages} {383} (\bibinfo {year} {1997})},\ \Eprint
  {http://arxiv.org/abs/hep-ex/9703005} {arXiv:hep-ex/9703005} \BibitemShut
  {NoStop}%
\bibitem [{\citenamefont {Airapetian}\ \emph {et~al.}(2007)\citenamefont
  {Airapetian} \emph {et~al.}}]{Airapetian:2007mh}%
  \BibitemOpen
  \bibfield  {author} {\bibinfo {author} {\bibfnamefont {A.}~\bibnamefont
  {Airapetian}} \emph {et~al.} (\bibinfo {collaboration} {HERMES
  Collaboration}),\ }\href {\doibase 10.1103/PhysRevD.75.012007} {\bibfield
  {journal} {\bibinfo  {journal} {Phys. Rev. D}\ }\textbf {\bibinfo {volume}
  {75}},\ \bibinfo {pages} {012007} (\bibinfo {year} {2007})},\ \Eprint
  {http://arxiv.org/abs/hep-ex/0609039} {arXiv:hep-ex/0609039} \BibitemShut
  {NoStop}%
\bibitem [{\citenamefont {Baum}\ \emph {et~al.}(1983)\citenamefont {Baum} \emph
  {et~al.}}]{Baum:1983ha}%
  \BibitemOpen
  \bibfield  {author} {\bibinfo {author} {\bibfnamefont {G.}~\bibnamefont
  {Baum}} \emph {et~al.},\ }\href {\doibase 10.1103/PhysRevLett.51.1135}
  {\bibfield  {journal} {\bibinfo  {journal} {Phys. Rev. Lett.}\ }\textbf
  {\bibinfo {volume} {51}},\ \bibinfo {pages} {1135} (\bibinfo {year}
  {1983})}\BibitemShut {NoStop}%
\bibitem [{\citenamefont {Prok}\ \emph {et~al.}(2014)\citenamefont {Prok} \emph
  {et~al.}}]{Prok:2014ltt}%
  \BibitemOpen
  \bibfield  {author} {\bibinfo {author} {\bibfnamefont {Y.}~\bibnamefont
  {Prok}} \emph {et~al.} (\bibinfo {collaboration} {CLAS}),\ }\href {\doibase
  10.1103/PhysRevC.90.025212} {\bibfield  {journal} {\bibinfo  {journal} {Phys.
  Rev. C}\ }\textbf {\bibinfo {volume} {90}},\ \bibinfo {pages} {025212}
  (\bibinfo {year} {2014})},\ \Eprint {http://arxiv.org/abs/1404.6231}
  {arXiv:1404.6231 [nucl-ex]} \BibitemShut {NoStop}%
\bibitem [{\citenamefont {Parno}\ \emph {et~al.}(2015)\citenamefont {Parno}
  \emph {et~al.}}]{Parno:2014xzb}%
  \BibitemOpen
  \bibfield  {author} {\bibinfo {author} {\bibfnamefont {D.~S.}\ \bibnamefont
  {Parno}} \emph {et~al.} (\bibinfo {collaboration} {Jefferson Lab Hall A}),\
  }\href {\doibase 10.1016/j.physletb.2015.03.067} {\bibfield  {journal}
  {\bibinfo  {journal} {Phys. Lett. B}\ }\textbf {\bibinfo {volume} {744}},\
  \bibinfo {pages} {309} (\bibinfo {year} {2015})},\ \Eprint
  {http://arxiv.org/abs/1406.1207} {arXiv:1406.1207 [nucl-ex]} \BibitemShut
  {NoStop}%
\bibitem [{\citenamefont {Guler}\ \emph {et~al.}(2015)\citenamefont {Guler}
  \emph {et~al.}}]{Guler:2015hsw}%
  \BibitemOpen
  \bibfield  {author} {\bibinfo {author} {\bibfnamefont {N.}~\bibnamefont
  {Guler}} \emph {et~al.} (\bibinfo {collaboration} {CLAS}),\ }\href {\doibase
  10.1103/PhysRevC.92.055201} {\bibfield  {journal} {\bibinfo  {journal} {Phys.
  Rev. C}\ }\textbf {\bibinfo {volume} {92}},\ \bibinfo {pages} {055201}
  (\bibinfo {year} {2015})},\ \Eprint {http://arxiv.org/abs/1505.07877}
  {arXiv:1505.07877 [nucl-ex]} \BibitemShut {NoStop}%
\bibitem [{\citenamefont {Fersch}\ \emph {et~al.}(2017)\citenamefont {Fersch}
  \emph {et~al.}}]{Fersch:2017qrq}%
  \BibitemOpen
  \bibfield  {author} {\bibinfo {author} {\bibfnamefont {R.}~\bibnamefont
  {Fersch}} \emph {et~al.} (\bibinfo {collaboration} {CLAS}),\ }\href {\doibase
  10.1103/PhysRevC.96.065208} {\bibfield  {journal} {\bibinfo  {journal} {Phys.
  Rev. C}\ }\textbf {\bibinfo {volume} {96}},\ \bibinfo {pages} {065208}
  (\bibinfo {year} {2017})},\ \Eprint {http://arxiv.org/abs/1706.10289}
  {arXiv:1706.10289 [nucl-ex]} \BibitemShut {NoStop}%
\bibitem [{\citenamefont {Kuraev}\ \emph {et~al.}(1977)\citenamefont {Kuraev},
  \citenamefont {Lipatov},\ and\ \citenamefont {Fadin}}]{Kuraev:1977fs}%
  \BibitemOpen
  \bibfield  {author} {\bibinfo {author} {\bibfnamefont {E.~A.}\ \bibnamefont
  {Kuraev}}, \bibinfo {author} {\bibfnamefont {L.~N.}\ \bibnamefont {Lipatov}},
  \ and\ \bibinfo {author} {\bibfnamefont {V.~S.}\ \bibnamefont {Fadin}},\
  }\href@noop {} {\bibfield  {journal} {\bibinfo  {journal} {Sov. Phys. JETP}\
  }\textbf {\bibinfo {volume} {45}},\ \bibinfo {pages} {199} (\bibinfo {year}
  {1977})}\BibitemShut {NoStop}%
\bibitem [{\citenamefont {Balitsky}\ and\ \citenamefont
  {Lipatov}(1978)}]{Balitsky:1978ic}%
  \BibitemOpen
  \bibfield  {author} {\bibinfo {author} {\bibfnamefont {I.~I.}\ \bibnamefont
  {Balitsky}}\ and\ \bibinfo {author} {\bibfnamefont {L.~N.}\ \bibnamefont
  {Lipatov}},\ }\href@noop {} {\bibfield  {journal} {\bibinfo  {journal} {Sov.
  J. Nucl. Phys.}\ }\textbf {\bibinfo {volume} {28}},\ \bibinfo {pages} {822}
  (\bibinfo {year} {1978})}\BibitemShut {NoStop}%
\bibitem [{\citenamefont {Jalilian-Marian}\ \emph
  {et~al.}(1998{\natexlab{a}})\citenamefont {Jalilian-Marian}, \citenamefont
  {Kovner},\ and\ \citenamefont {Weigert}}]{JalilianMarian:1997dw}%
  \BibitemOpen
  \bibfield  {author} {\bibinfo {author} {\bibfnamefont {J.}~\bibnamefont
  {Jalilian-Marian}}, \bibinfo {author} {\bibfnamefont {A.}~\bibnamefont
  {Kovner}}, \ and\ \bibinfo {author} {\bibfnamefont {H.}~\bibnamefont
  {Weigert}},\ }\href {\doibase 10.1103/PhysRevD.59.014015} {\bibfield
  {journal} {\bibinfo  {journal} {Phys. Rev. D}\ }\textbf {\bibinfo {volume}
  {59}},\ \bibinfo {pages} {014015} (\bibinfo {year} {1998}{\natexlab{a}})},\
  \Eprint {http://arxiv.org/abs/hep-ph/9709432} {arXiv:hep-ph/9709432}
  \BibitemShut {NoStop}%
\bibitem [{\citenamefont {Jalilian-Marian}\ \emph
  {et~al.}(1998{\natexlab{b}})\citenamefont {Jalilian-Marian}, \citenamefont
  {Kovner}, \citenamefont {Leonidov},\ and\ \citenamefont
  {Weigert}}]{JalilianMarian:1997gr}%
  \BibitemOpen
  \bibfield  {author} {\bibinfo {author} {\bibfnamefont {J.}~\bibnamefont
  {Jalilian-Marian}}, \bibinfo {author} {\bibfnamefont {A.}~\bibnamefont
  {Kovner}}, \bibinfo {author} {\bibfnamefont {A.}~\bibnamefont {Leonidov}}, \
  and\ \bibinfo {author} {\bibfnamefont {H.}~\bibnamefont {Weigert}},\ }\href
  {\doibase 10.1103/PhysRevD.59.014014} {\bibfield  {journal} {\bibinfo
  {journal} {Phys. Rev. D}\ }\textbf {\bibinfo {volume} {59}},\ \bibinfo
  {pages} {014014} (\bibinfo {year} {1998}{\natexlab{b}})},\ \Eprint
  {http://arxiv.org/abs/hep-ph/9706377} {arXiv:hep-ph/9706377} \BibitemShut
  {NoStop}%
\bibitem [{\citenamefont {Weigert}(2002)}]{Weigert:2000gi}%
  \BibitemOpen
  \bibfield  {author} {\bibinfo {author} {\bibfnamefont {H.}~\bibnamefont
  {Weigert}},\ }\href {\doibase 10.1016/S0375-9474(01)01668-2} {\bibfield
  {journal} {\bibinfo  {journal} {Nucl. Phys.}\ }\textbf {\bibinfo {volume}
  {A703}},\ \bibinfo {pages} {823} (\bibinfo {year} {2002})},\ \Eprint
  {http://arxiv.org/abs/hep-ph/0004044} {arXiv:hep-ph/0004044} \BibitemShut
  {NoStop}%
\bibitem [{\citenamefont {Iancu}\ \emph
  {et~al.}(2001{\natexlab{a}})\citenamefont {Iancu}, \citenamefont {Leonidov},\
  and\ \citenamefont {McLerran}}]{Iancu:2001ad}%
  \BibitemOpen
  \bibfield  {author} {\bibinfo {author} {\bibfnamefont {E.}~\bibnamefont
  {Iancu}}, \bibinfo {author} {\bibfnamefont {A.}~\bibnamefont {Leonidov}}, \
  and\ \bibinfo {author} {\bibfnamefont {L.~D.}\ \bibnamefont {McLerran}},\
  }\href {\doibase 10.1016/S0370-2693(01)00524-X} {\bibfield  {journal}
  {\bibinfo  {journal} {Phys. Lett. B}\ }\textbf {\bibinfo {volume} {510}},\
  \bibinfo {pages} {133} (\bibinfo {year} {2001}{\natexlab{a}})},\ \Eprint
  {http://arxiv.org/abs/hep-ph/0102009} {arXiv:hep-ph/0102009} \BibitemShut
  {NoStop}%
\bibitem [{\citenamefont {Iancu}\ \emph
  {et~al.}(2001{\natexlab{b}})\citenamefont {Iancu}, \citenamefont {Leonidov},\
  and\ \citenamefont {McLerran}}]{Iancu:2000hn}%
  \BibitemOpen
  \bibfield  {author} {\bibinfo {author} {\bibfnamefont {E.}~\bibnamefont
  {Iancu}}, \bibinfo {author} {\bibfnamefont {A.}~\bibnamefont {Leonidov}}, \
  and\ \bibinfo {author} {\bibfnamefont {L.~D.}\ \bibnamefont {McLerran}},\
  }\href {\doibase 10.1016/S0375-9474(01)00642-X} {\bibfield  {journal}
  {\bibinfo  {journal} {Nucl. Phys.}\ }\textbf {\bibinfo {volume} {A692}},\
  \bibinfo {pages} {583} (\bibinfo {year} {2001}{\natexlab{b}})},\ \Eprint
  {http://arxiv.org/abs/hep-ph/0011241} {arXiv:hep-ph/0011241} \BibitemShut
  {NoStop}%
\bibitem [{\citenamefont {Ferreiro}\ \emph {et~al.}(2002)\citenamefont
  {Ferreiro}, \citenamefont {Iancu}, \citenamefont {Leonidov},\ and\
  \citenamefont {McLerran}}]{Ferreiro:2001qy}%
  \BibitemOpen
  \bibfield  {author} {\bibinfo {author} {\bibfnamefont {E.}~\bibnamefont
  {Ferreiro}}, \bibinfo {author} {\bibfnamefont {E.}~\bibnamefont {Iancu}},
  \bibinfo {author} {\bibfnamefont {A.}~\bibnamefont {Leonidov}}, \ and\
  \bibinfo {author} {\bibfnamefont {L.}~\bibnamefont {McLerran}},\ }\href
  {\doibase 10.1016/S0375-9474(01)01329-X} {\bibfield  {journal} {\bibinfo
  {journal} {Nucl. Phys.}\ }\textbf {\bibinfo {volume} {A703}},\ \bibinfo
  {pages} {489} (\bibinfo {year} {2002})},\ \Eprint
  {http://arxiv.org/abs/hep-ph/0109115} {arXiv:hep-ph/0109115} \BibitemShut
  {NoStop}%
\bibitem [{\citenamefont {Kharzeev}\ \emph {et~al.}(2003)\citenamefont
  {Kharzeev}, \citenamefont {Kovchegov},\ and\ \citenamefont
  {Tuchin}}]{Kharzeev:2003wz}%
  \BibitemOpen
  \bibfield  {author} {\bibinfo {author} {\bibfnamefont {D.}~\bibnamefont
  {Kharzeev}}, \bibinfo {author} {\bibfnamefont {Y.~V.}\ \bibnamefont
  {Kovchegov}}, \ and\ \bibinfo {author} {\bibfnamefont {K.}~\bibnamefont
  {Tuchin}},\ }\href {\doibase 10.1103/PhysRevD.68.094013} {\bibfield
  {journal} {\bibinfo  {journal} {Phys. Rev. D}\ }\textbf {\bibinfo {volume}
  {68}},\ \bibinfo {pages} {094013} (\bibinfo {year} {2003})},\ \Eprint
  {http://arxiv.org/abs/hep-ph/0307037} {arXiv:hep-ph/0307037} \BibitemShut
  {NoStop}%
\bibitem [{\citenamefont {Albacete}\ \emph {et~al.}(2004)\citenamefont
  {Albacete}, \citenamefont {Armesto}, \citenamefont {Kovner}, \citenamefont
  {Salgado},\ and\ \citenamefont {Wiedemann}}]{Albacete:2003iq}%
  \BibitemOpen
  \bibfield  {author} {\bibinfo {author} {\bibfnamefont {J.~L.}\ \bibnamefont
  {Albacete}}, \bibinfo {author} {\bibfnamefont {N.}~\bibnamefont {Armesto}},
  \bibinfo {author} {\bibfnamefont {A.}~\bibnamefont {Kovner}}, \bibinfo
  {author} {\bibfnamefont {C.~A.}\ \bibnamefont {Salgado}}, \ and\ \bibinfo
  {author} {\bibfnamefont {U.~A.}\ \bibnamefont {Wiedemann}},\ }\href {\doibase
  10.1103/PhysRevLett.92.082001} {\bibfield  {journal} {\bibinfo  {journal}
  {Phys. Rev. Lett.}\ }\textbf {\bibinfo {volume} {92}},\ \bibinfo {pages}
  {082001} (\bibinfo {year} {2004})},\ \Eprint
  {http://arxiv.org/abs/hep-ph/0307179} {arXiv:hep-ph/0307179} \BibitemShut
  {NoStop}%
\bibitem [{\citenamefont {Kharzeev}\ \emph {et~al.}(2004)\citenamefont
  {Kharzeev}, \citenamefont {Kovchegov},\ and\ \citenamefont
  {Tuchin}}]{Kharzeev:2004yx}%
  \BibitemOpen
  \bibfield  {author} {\bibinfo {author} {\bibfnamefont {D.}~\bibnamefont
  {Kharzeev}}, \bibinfo {author} {\bibfnamefont {Y.~V.}\ \bibnamefont
  {Kovchegov}}, \ and\ \bibinfo {author} {\bibfnamefont {K.}~\bibnamefont
  {Tuchin}},\ }\href {\doibase 10.1016/j.physletb.2004.08.034} {\bibfield
  {journal} {\bibinfo  {journal} {Phys. Lett. B}\ }\textbf {\bibinfo {volume}
  {599}},\ \bibinfo {pages} {23} (\bibinfo {year} {2004})},\ \Eprint
  {http://arxiv.org/abs/hep-ph/0405045} {arXiv:hep-ph/0405045} \BibitemShut
  {NoStop}%
\bibitem [{\citenamefont {Albacete}\ \emph {et~al.}(2005)\citenamefont
  {Albacete}, \citenamefont {Armesto}, \citenamefont {Milhano}, \citenamefont
  {Salgado},\ and\ \citenamefont {Wiedemann}}]{Albacete:2004gw}%
  \BibitemOpen
  \bibfield  {author} {\bibinfo {author} {\bibfnamefont {J.}~\bibnamefont
  {Albacete}}, \bibinfo {author} {\bibfnamefont {N.}~\bibnamefont {Armesto}},
  \bibinfo {author} {\bibfnamefont {J.}~\bibnamefont {Milhano}}, \bibinfo
  {author} {\bibfnamefont {C.}~\bibnamefont {Salgado}}, \ and\ \bibinfo
  {author} {\bibfnamefont {U.}~\bibnamefont {Wiedemann}},\ }\href {\doibase
  10.1103/PhysRevD.71.014003} {\bibfield  {journal} {\bibinfo  {journal} {Phys.
  Rev. D}\ }\textbf {\bibinfo {volume} {71}},\ \bibinfo {pages} {014003}
  (\bibinfo {year} {2005})},\ \Eprint {http://arxiv.org/abs/hep-ph/0408216}
  {arXiv:hep-ph/0408216} \BibitemShut {NoStop}%
\bibitem [{\citenamefont {de~Florian}\ \emph {et~al.}(2014)\citenamefont
  {de~Florian}, \citenamefont {Sassot}, \citenamefont {Stratmann},\ and\
  \citenamefont {Vogelsang}}]{deFlorian:2014yva}%
  \BibitemOpen
  \bibfield  {author} {\bibinfo {author} {\bibfnamefont {D.}~\bibnamefont
  {de~Florian}}, \bibinfo {author} {\bibfnamefont {R.}~\bibnamefont {Sassot}},
  \bibinfo {author} {\bibfnamefont {M.}~\bibnamefont {Stratmann}}, \ and\
  \bibinfo {author} {\bibfnamefont {W.}~\bibnamefont {Vogelsang}},\ }\href
  {\doibase 10.1103/PhysRevLett.113.012001} {\bibfield  {journal} {\bibinfo
  {journal} {Phys. Rev. Lett.}\ }\textbf {\bibinfo {volume} {113}},\ \bibinfo
  {pages} {012001} (\bibinfo {year} {2014})},\ \Eprint
  {http://arxiv.org/abs/1404.4293} {arXiv:1404.4293 [hep-ph]} \BibitemShut
  {NoStop}%
\bibitem [{\citenamefont {De~Florian}\ \emph {et~al.}(2019)\citenamefont
  {De~Florian}, \citenamefont {Lucero}, \citenamefont {Sassot}, \citenamefont
  {Stratmann},\ and\ \citenamefont {Vogelsang}}]{deFlorian:2019zkl}%
  \BibitemOpen
  \bibfield  {author} {\bibinfo {author} {\bibfnamefont {D.}~\bibnamefont
  {De~Florian}}, \bibinfo {author} {\bibfnamefont {G.~A.}\ \bibnamefont
  {Lucero}}, \bibinfo {author} {\bibfnamefont {R.}~\bibnamefont {Sassot}},
  \bibinfo {author} {\bibfnamefont {M.}~\bibnamefont {Stratmann}}, \ and\
  \bibinfo {author} {\bibfnamefont {W.}~\bibnamefont {Vogelsang}},\ }\href
  {\doibase 10.1103/PhysRevD.100.114027} {\bibfield  {journal} {\bibinfo
  {journal} {Phys. Rev. D}\ }\textbf {\bibinfo {volume} {100}},\ \bibinfo
  {pages} {114027} (\bibinfo {year} {2019})},\ \Eprint
  {http://arxiv.org/abs/1902.10548} {arXiv:1902.10548 [hep-ph]} \BibitemShut
  {NoStop}%
\bibitem [{\citenamefont {Borsa}\ \emph {et~al.}(2020)\citenamefont {Borsa},
  \citenamefont {Lucero}, \citenamefont {Sassot}, \citenamefont {Aschenauer},\
  and\ \citenamefont {Nunes}}]{Aschenauer:2020pdk}%
  \BibitemOpen
  \bibfield  {author} {\bibinfo {author} {\bibfnamefont {I.}~\bibnamefont
  {Borsa}}, \bibinfo {author} {\bibfnamefont {G.}~\bibnamefont {Lucero}},
  \bibinfo {author} {\bibfnamefont {R.}~\bibnamefont {Sassot}}, \bibinfo
  {author} {\bibfnamefont {E.~C.}\ \bibnamefont {Aschenauer}}, \ and\ \bibinfo
  {author} {\bibfnamefont {A.~S.}\ \bibnamefont {Nunes}},\ }\href {\doibase
  10.1103/PhysRevD.102.094018} {\bibfield  {journal} {\bibinfo  {journal}
  {Phys. Rev. D}\ }\textbf {\bibinfo {volume} {102}},\ \bibinfo {pages}
  {094018} (\bibinfo {year} {2020})},\ \Eprint
  {http://arxiv.org/abs/2007.08300} {arXiv:2007.08300 [hep-ph]} \BibitemShut
  {NoStop}%
\bibitem [{\citenamefont {Kirschner}(1986)}]{Kirschner:1985cb}%
  \BibitemOpen
  \bibfield  {author} {\bibinfo {author} {\bibfnamefont {R.}~\bibnamefont
  {Kirschner}},\ }\href {\doibase 10.1007/BF01559604} {\bibfield  {journal}
  {\bibinfo  {journal} {Z. Phys. C}\ }\textbf {\bibinfo {volume} {31}},\
  \bibinfo {pages} {135} (\bibinfo {year} {1986})}\BibitemShut {NoStop}%
\bibitem [{\citenamefont {Kirschner}(1995{\natexlab{a}})}]{Kirschner:1994vc}%
  \BibitemOpen
  \bibfield  {author} {\bibinfo {author} {\bibfnamefont {R.}~\bibnamefont
  {Kirschner}},\ }\href {\doibase 10.1007/BF01624588} {\bibfield  {journal}
  {\bibinfo  {journal} {Z. Phys. C}\ }\textbf {\bibinfo {volume} {67}},\
  \bibinfo {pages} {459} (\bibinfo {year} {1995}{\natexlab{a}})},\ \Eprint
  {http://arxiv.org/abs/hep-th/9404158} {arXiv:hep-th/9404158} \BibitemShut
  {NoStop}%
\bibitem [{\citenamefont {Kirschner}(1995{\natexlab{b}})}]{Kirschner:1994rq}%
  \BibitemOpen
  \bibfield  {author} {\bibinfo {author} {\bibfnamefont {R.}~\bibnamefont
  {Kirschner}},\ }\href {\doibase 10.1007/BF01556138} {\bibfield  {journal}
  {\bibinfo  {journal} {Z. Phys. C}\ }\textbf {\bibinfo {volume} {65}},\
  \bibinfo {pages} {505} (\bibinfo {year} {1995}{\natexlab{b}})},\ \Eprint
  {http://arxiv.org/abs/hep-th/9407085} {arXiv:hep-th/9407085} \BibitemShut
  {NoStop}%
\bibitem [{\citenamefont {Griffiths}\ and\ \citenamefont
  {Ross}(2000)}]{Griffiths:1999dj}%
  \BibitemOpen
  \bibfield  {author} {\bibinfo {author} {\bibfnamefont {S.}~\bibnamefont
  {Griffiths}}\ and\ \bibinfo {author} {\bibfnamefont {D.~A.}\ \bibnamefont
  {Ross}},\ }\href {\doibase 10.1007/s100529900240} {\bibfield  {journal}
  {\bibinfo  {journal} {Eur. Phys. J. C}\ }\textbf {\bibinfo {volume} {12}},\
  \bibinfo {pages} {277} (\bibinfo {year} {2000})},\ \Eprint
  {http://arxiv.org/abs/hep-ph/9906550} {arXiv:hep-ph/9906550} \BibitemShut
  {NoStop}%
\bibitem [{\citenamefont {Ermolaev}\ \emph {et~al.}(1996)\citenamefont
  {Ermolaev}, \citenamefont {Manaenkov},\ and\ \citenamefont
  {Ryskin}}]{Ermolaev:1995fx}%
  \BibitemOpen
  \bibfield  {author} {\bibinfo {author} {\bibfnamefont {B.~I.}\ \bibnamefont
  {Ermolaev}}, \bibinfo {author} {\bibfnamefont {S.~I.}\ \bibnamefont
  {Manaenkov}}, \ and\ \bibinfo {author} {\bibfnamefont {M.~G.}\ \bibnamefont
  {Ryskin}},\ }\href {\doibase 10.1007/s002880050026} {\bibfield  {journal}
  {\bibinfo  {journal} {Z. Phys. C}\ }\textbf {\bibinfo {volume} {69}},\
  \bibinfo {pages} {259} (\bibinfo {year} {1996})},\ \Eprint
  {http://arxiv.org/abs/hep-ph/9502262} {arXiv:hep-ph/9502262} \BibitemShut
  {NoStop}%
\bibitem [{\citenamefont {Bartels}\ and\ \citenamefont
  {Lublinsky}(2003)}]{Bartels:2003dj}%
  \BibitemOpen
  \bibfield  {author} {\bibinfo {author} {\bibfnamefont {J.}~\bibnamefont
  {Bartels}}\ and\ \bibinfo {author} {\bibfnamefont {M.}~\bibnamefont
  {Lublinsky}},\ }\href {\doibase 10.1088/1126-6708/2003/09/076} {\bibfield
  {journal} {\bibinfo  {journal} {JHEP}\ }\textbf {\bibinfo {volume} {09}},\
  \bibinfo {pages} {076} (\bibinfo {year} {2003})},\ \Eprint
  {http://arxiv.org/abs/hep-ph/0308181} {arXiv:hep-ph/0308181} \BibitemShut
  {NoStop}%
\bibitem [{\citenamefont {Donnachie}\ and\ \citenamefont
  {Landshoff}(1992)}]{Donnachie:1992ny}%
  \BibitemOpen
  \bibfield  {author} {\bibinfo {author} {\bibfnamefont {A.}~\bibnamefont
  {Donnachie}}\ and\ \bibinfo {author} {\bibfnamefont {P.~V.}\ \bibnamefont
  {Landshoff}},\ }\href {\doibase 10.1016/0370-2693(92)90832-O} {\bibfield
  {journal} {\bibinfo  {journal} {Phys. Lett. B}\ }\textbf {\bibinfo {volume}
  {296}},\ \bibinfo {pages} {227} (\bibinfo {year} {1992})},\ \Eprint
  {http://arxiv.org/abs/hep-ph/9209205} {arXiv:hep-ph/9209205} \BibitemShut
  {NoStop}%
\bibitem [{\citenamefont {Itakura}\ \emph {et~al.}(2004)\citenamefont
  {Itakura}, \citenamefont {Kovchegov}, \citenamefont {McLerran},\ and\
  \citenamefont {Teaney}}]{Itakura:2003jp}%
  \BibitemOpen
  \bibfield  {author} {\bibinfo {author} {\bibfnamefont {K.}~\bibnamefont
  {Itakura}}, \bibinfo {author} {\bibfnamefont {Y.~V.}\ \bibnamefont
  {Kovchegov}}, \bibinfo {author} {\bibfnamefont {L.}~\bibnamefont {McLerran}},
  \ and\ \bibinfo {author} {\bibfnamefont {D.}~\bibnamefont {Teaney}},\ }\href
  {\doibase 10.1016/j.nuclphysa.2003.10.016} {\bibfield  {journal} {\bibinfo
  {journal} {Nucl. Phys.}\ }\textbf {\bibinfo {volume} {A730}},\ \bibinfo
  {pages} {160} (\bibinfo {year} {2004})},\ \Eprint
  {http://arxiv.org/abs/hep-ph/0305332} {arXiv:hep-ph/0305332} \BibitemShut
  {NoStop}%
\bibitem [{\citenamefont {Abdul~Khalek}\ \emph {et~al.}(2021)\citenamefont
  {Abdul~Khalek} \emph {et~al.}}]{AbdulKhalek:2021gbh}%
  \BibitemOpen
  \bibfield  {author} {\bibinfo {author} {\bibfnamefont {R.}~\bibnamefont
  {Abdul~Khalek}} \emph {et~al.},\ }\href@noop {} {\  (\bibinfo {year}
  {2021})},\ \Eprint {http://arxiv.org/abs/2103.05419} {arXiv:2103.05419
  [physics.ins-det]} \BibitemShut {NoStop}%
\bibitem [{\citenamefont {Kovchegov}\ \emph {et~al.}(2021)\citenamefont
  {Kovchegov}, \citenamefont {Tarasov},\ and\ \citenamefont
  {Tawabutr}}]{Kovchegov:2021lvz}%
  \BibitemOpen
  \bibfield  {author} {\bibinfo {author} {\bibfnamefont {Y.~V.}\ \bibnamefont
  {Kovchegov}}, \bibinfo {author} {\bibfnamefont {A.}~\bibnamefont {Tarasov}},
  \ and\ \bibinfo {author} {\bibfnamefont {Y.}~\bibnamefont {Tawabutr}},\
  }\href@noop {} {\  (\bibinfo {year} {2021})},\ \Eprint
  {http://arxiv.org/abs/2104.11765} {arXiv:2104.11765 [hep-ph]} \BibitemShut
  {NoStop}%
\end{thebibliography}
\end{document}